\documentclass[a4paper,fleqn,usenatbib]{mnras}
\usepackage{graphicx}
\usepackage{amssymb}
\usepackage{amsmath}
\usepackage{ae,aecompl}
\usepackage{natbib,times}
\usepackage{lscape}
\newcommand{\mc}{\multicolumn}

\begin{document}
%%%%%%%%%%%%%%%%%%%%%%%%%%%%%%%%%%%%%%%%%
\title[Photo-$z$ and clusters from HSC$\times$unWISE]
      {Photometric redshifts for galaxies in the Subaru Hyper Suprime-Cam
        and unWISE and a catalogue of identified clusters of galaxies}

\author[Wen \& Han]
{Z. L. Wen$^{1,2}$\thanks{E-mail: zhonglue@nao.cas.cn}
  and J. L. Han$^{1,2,3}$
\\
1. National Astronomical Observatories, Chinese Academy of Sciences, 
20A Datun Road, Chaoyang District, Beijing 100101, China\\
2. CAS Key Laboratory of FAST, NAOC, Chinese Academy of Sciences,
           Beijing 100101, China \\
3. School of Astronomy, University of Chinese Academy of Sciences,
           Beijing 100049, China 
}

\date{Accepted XXX. Received YYY; in original form ZZZ}

\label{firstpage}
\pagerange{\pageref{firstpage}--\pageref{lastpage}}
\maketitle

%%%%%%%%%%%%%%%%%%%%%%%%%%%%%%%%%%%%%%%%%%%%%%%%%%%%%%%%%%%%%%%%%%%%%%%%%%%%%

\begin{abstract}
We first present a catalogue of photometric redshifts for 14.68
million galaxies derived from the 7-band photometric data of
Hyper Suprime-Cam Subaru Strategic Program and the Wide-field Infrared
Survey Explorer using the nearest-neighbour algorithm. The
redshift uncertainty is about 0.024 for galaxies of $z\le0.7$, and
steadily increases with redshift to about 0.11 at $z\sim2$.
From such a large data set, we identify 21,661 clusters of galaxies,
among which 5537 clusters have redshifts $z>1$ and 642 clusters have
$z>1.5$, significantly enlarging the high redshift sample of galaxy
clusters. Cluster richness and mass are estimated, and these clusters
have an equivalent mass of $M_{500} \ge 0.7\times10^{14}M_{\odot}$. We
find that the stellar mass of the brightest cluster galaxies (BCGs) in
each richness bin does not significantly evolve with redshift. The
fraction of star-forming BCGs increases with redshift, but does not
depend on cluster mass.

\end{abstract}


%%%%%%%%%%%%%%%%%%%%%%%%%%%%%%%%%%%%%%%%%%%%%%%%%%%%%%%%%%%%%%%%%%%%%%%
\section{Introduction}

According to the hierarchical scenario \citep{pee80}, clusters of
galaxies were formed at knots of cosmic web and grow through accretion
and merging of smaller structures \citep{cwj+99}. Galaxy clusters
therefore are one of important tracers of the large scale structure
(LSS) of the Universe \citep{aem11}.
Cosmological parameters can be constrained from the evolution of
cluster number count and gas mass fraction \citep{ars+08,vkb+09}, even
constrained by a very few extremely massive clusters at high redshifts
of $z>1$ \citep{jrf+09,hjv11,bgs+12}.
The properties of galaxy component in clusters show significant
evolution with redshift. Cluster galaxies in the local universe are
dominated by massive early-type galaxies \citep{roo69,bts87}. Clusters
at higher redshifts possess a higher fraction of star-forming
galaxies, which is known as the Butcher--Oemler effect
\citep{bo78,bo84}. Galaxy population in some clusters of $z>2$ is
dominated by galaxies with a very high star formation rate
\citep[e.g.][]{wed+16,sgs+19}.
Specially, the brightest cluster galaxies (BCGs) are the most massive
galaxies. Their properties are related to the mass and dynamical state of
host clusters \citep{kvc+15,wh15,eff+19}. The stellar population in
BCGs was generally formed at redshifts $z>2$ and evolved passively
afterwards \citep{ses+08,wad+08,wh11}.

Systematic searches for galaxy clusters from multiwavelength survey
data have been carried out over a few decades since \citet{abe58} and
\citet{aco89}. A large number of optical clusters have been found from
the Sloan Digital Sky Survey (SDSS), most of which have a redshift of
$z\lesssim0.7$
\citep{kma+07b,whl09,hmk+10,spd+11,whl12,ogu14,rrb+14,bsp+18}. In
addition, a few thousand clusters were found from the ROentgen
SATellite all sky X-ray survey \citep{pap+11,bcc+13,bcr+17}
and Planck millimeter survey \citep{plancksz16}.
At higher redshifts, some galaxy clusters have been found from 
deep field data covering a small area
\citep[e.g.][]{gy05,vcb+06,ebg+08,wh11}. Recently, large cluster
samples at $z>0.7$ were identified from the combination of optical and
infrared data, for example, 1959 massive clusters of $0.7<z<1$ from
the photometric data of the SDSS and Wide-field Infrared Survey
Explorer (WISE) by \citet{wh18}, 2433 massive clusters of $0.7<z<1.5$
from the WISE data supplemented with the data of Pan-STARRS and
SuperCOSMOS by \citet{ggb+19}. Via the Sunyaev-Zel'dovich (SZ) effect,
more than one hundred massive clusters of $z>0.7$ have been previously
detected from the sky surveys of cosmic microwave background by
Atacama Cosmology Telescope (ACT) and South Pole Telescope 
\citep{rsb+13,hhs+18}. Up to now, only a few tens of clusters at
$z>1.5$ have been identified or confirmed individually from optical,
X-ray and SZ data \citep[e.g.][]{pmw+10,tsj15,hhs+18}.

Usually, optical data supplemented with infrared data are efficient to
reveal high-redshift clusters. Recent public survey data provide a new
opportunity for finding high-redshift galaxy
clusters. Hyper Suprime-Cam Subaru Strategic Program (HSC-SSP) is a
large area optical survey down to a limit of $i=26$ mag, which is deep
enough to detect massive galaxy of $z>2$
\citep{hscssp18}. 
The WISE is an all sky survey in four mid-infrared bands
\citep{wem+10}. The recent public unWISE catalogue of the WISE reaches
a limit of ${\rm W1}=17.8$ mag and contains many high-redshift
massive galaxies \citep{smg19}.

In this paper, we first present a catalogue of photometric redshifts
for galaxies in the cross-matched catalogue of the HSC-SSP and the
unWISE, containing improved photometric redshifts of galaxies at
$1.5<z<2$. Subsequently, a catalogue of 21,661 galaxy clusters of
$0.1<z\lesssim2$ are identified, which significantly enlarges the
number of clusters at $z>1$.
In Section 2, we first describe the galaxy data and the estimates of
galaxy photometric redshift and stellar mass. In Section 3, we
describe the cluster identification procedures and the identified
galaxy clusters. In Section 4, we study the evolution of BCGs,
including their stellar mass and star formation.  A summary is
presented in Section 5.

Throughout this paper, we assume a flat Lambda cold dark matter
cosmology taking $H_0=70$ km~s$^{-1}$ Mpc$^{-1}$, $\Omega_m=0.3$ and
$\Omega_{\Lambda}=0.7$.

\section{Photometric redshifts and stellar masses of galaxies}

Based on large data set of photometric surveys, the photometric
redshift of galaxies is a fundamental parameter for many further
studies, such as galaxy evolution, structure formation and cluster
identification. We cross-match the galaxy catalogue of the HSC-SSP and
the source catalogue of the unWISE to get a 7-band spectral energy
distribution (SED) for the common galaxies in the two catalogues. The
photometric redshift of a galaxy can thus be estimated by the
comparison with a spectroscopic training sample in colour space. In
addition, the stellar mass of galaxies is another important parameter
in the algorithm for cluster identification, and can be estimated from
the photometric data.

\subsection{Galaxy data}

The HSC-SSP\footnote{https://hsc.mtk.nao.ac.jp/ssp/} carries out the
optical photometric survey in five broad bands ($grizy$) and three
narrow bands (NB387, NB816 and NB921) \citep{hscssp18}. It is
currently the deepest large-area optical survey, and consists of three
layers of survey depth. The Wide survey aims to observe a large area
of the sky covering 1400 deg$^2$ in the five broad bands, reaching a
5$\sigma$ limit of $i\sim26$ mag for point sources. The Deep survey
aims to observe four separate fields (XMM-LSS, Extended-COSMOS,
ELAIS-N1 and DEEP2-F3) covering about 27 deg$^2$ in the five broad
bands and the three narrow bands, reaching a limit of $i\sim27$
mag. The UltraDeep survey aims to observe two separate fields (COSMOS
and SXDS) covering about 3.5 deg$^2$ in the five broad bands and the
three narrow bands, reaching a limit of $i\sim28$ mag. The latest
HSC-SSP second data release (DR2) is now publicly available for the
Wide survey covering 924, 1022, 796, 905 and 924 deg$^2$ in the $g$,
$r$, $i$, $z$ and $y$ bands, respectively, and covering about 35
deg$^2$ for the Deep$+$UltraDeep survey \citep{hscdr2}. The
Deep$+$UltraDeep field is relatively small and mostly (except the
ELAIS-N1 field) overlapped by the Wide field. We here only consider
the five-band photometric HSC-SSP data in the Wide field and the
ELAIS-N1 field down to $i=26$ with a total sky area of $\sim 800$
deg$^2$. From the HSC-SSP DR2, we get 168 million galaxies observed at
least in the $r$, $i$ and $z$ bands with the flags of isprimary='True'
(i.e. a source has no children) and
i\_pixelflags\_saturatedcenter='False' (i.e. the source centre has no
saturated pixel in the $i$ band).

The WISE
survey\footnote{http://irsa.ipac.caltech.edu/Missions/wise.html/}
observed the whole sky in four mid-infrared bands \citep{wem+10}:
${\rm W1}$ (3.4 $\mu$m), ${\rm W2}$ (4.6 $\mu$m), ${\rm W3}$ (12
$\mu$m), and ${\rm W4}$ (22 $\mu$m). The WISE primary mission finished
the all sky survey in 2011 and published the photometric data, known as AllWISE
catalogue, in the four bands with 5$\sigma$ limits of 17.1, 15.7,
11.5, and 7.7 mag (Vega system), respectively, for point sources
\citep{allwise13}. Afterwards, an asteroid-characterizing extension
mission, NEOWISE, was carried out to continually survey the whole sky
in the ${\rm W1}$ and ${\rm W2}$ bands \citep{mbc+14}.
The unWISE catalogue\footnote{http://unwise.me/} was presented for about 2 billion
sources from the unblurred coadds of five-year NEOWISE single-exposure
images \citep{lang14,smg19}. It reaches a limit of 0.7 mag fainter than
the AllWISE catalogue and is deep enough to study massive galaxies to
$z\sim2$. We get 30 million unWISE sources with ${\rm W1}$-band data
available in the sky coverage of the HSC-SSP DR2. This number is much
smaller than that of the HSC-SSP galaxies because the unWISE catalogue
tends to include massive galaxies with a high mid-infrared flux.  We
convert the Vega magnitudes to AB magnitudes for the WISE sources by
adding the offsets of 2.699 and 3.339 for ${\rm W1}$ and ${\rm W2}$
magnitudes, respectively \citep{jcm+11}.

We perform one-to-one matching between the HSC-SSP catalogue and the
unWISE catalogue for the common galaxies, to form a catalogue of
HSC-SSP$\times$unWISE galaxies with 7-band photometric data. The
unWISE sources are cross-matched with each HSC-SSP galaxy using a
matching radius of 2 arcsec \citep{bpj+16,ske+19}. For the cases that
multiple matches are found, the closest source is adopted. Then,
around each matched unWISE source, the HSC-SSP galaxies are searched
within the offset of 2 arcsec and the closest one is adopted for the
multiple matches. Finally, we get 14.68 million HSC-SSP$\times$unWISE
galaxies.

\begin{figure}
\centering \includegraphics[width = 0.4\textwidth]{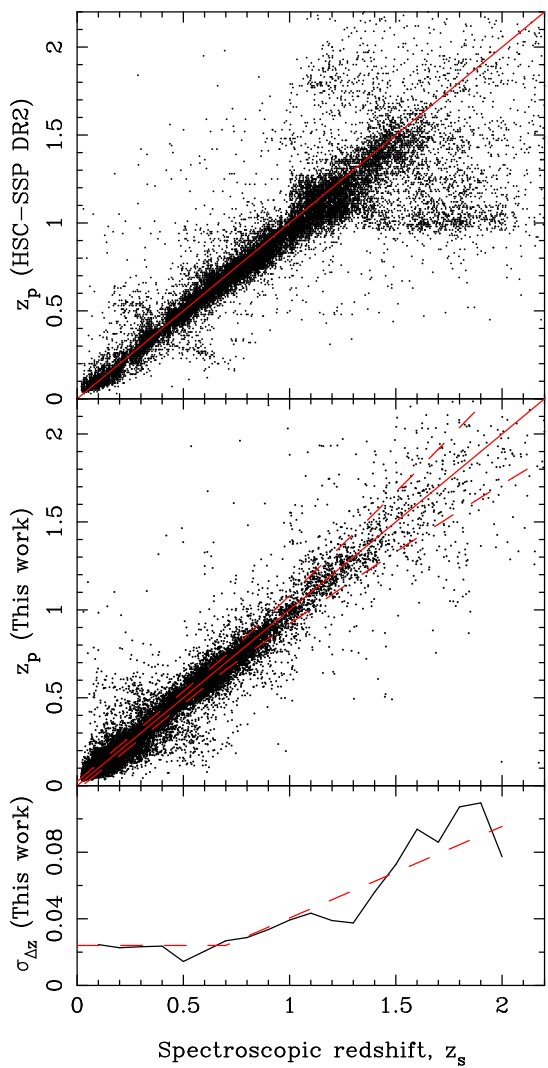}
\caption{Comparison between photometric redshifts and spectroscopic
  redshifts for the HSC-SSP DR2 estimates (upper panel) and the
  estimates of this work (middle panel). Lower: the uncertainty of
  photometric redshifts of this work. The dashed lines in the
  middle panel indicate the deviation of $1\,\sigma_{\Delta z}$, which
  is shown by the dashed line in the lower panel.}
\label{photoz}
\end{figure}

\begin{figure}
\centering \includegraphics[width = 0.4\textwidth]{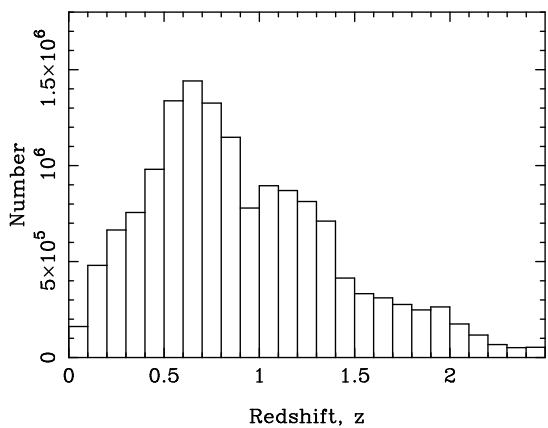}
\caption{Distribution of photometric redshifts for 14.68 million
  HSC-SSP$\times$unWISE galaxies.}
\label{histgz}
\end{figure}

\begin{figure}
\includegraphics[width = 0.47\textwidth]{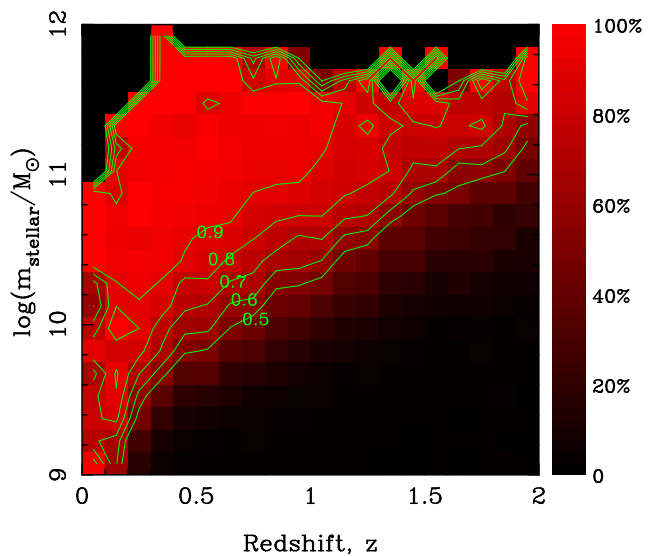}
\caption{Completeness of galaxies in the plane of redshift and galaxy
  stellar mass by comparison with the COSMOS2015 data. The level of
  the contour decreases from 90\% to 50\% by an interval of 10\%.}
\label{galcomplete}
\end{figure}

\subsection{Photometric redshifts of galaxies}

The 4000~\AA~break and 1.6 $\mu$m bump are distinct features to
determine photometric redshifts of galaxies. The 7-band photometric
data of the HSC-SSP$\times$unWISE galaxies can be used for the
estimation of photometric redshifts up to $z\sim2$. Usually, the
template fitting and the empirical methods are two main approaches for
the estimation of photometric redshifts. The former is applied by
fitting the observed SED with a set of template SEDs in various
parameter spaces to determine the most probable redshift for a galaxy
\citep[e.g.][]{bmp00,iam+06,bvc08,tan15}. The latter is requiring a
representative training sample with known spectroscopic redshifts. The
photometric redshift is determined from the empirical relations
between the photometric observables and the spectroscopic redshift
\citep[e.g.][]{cl04,hy14,cb14}. Here, we estimate photometric
redshifts of the HSC-SSP$\times$unWISE galaxies by an empirical
method, the nearest-neighbour algorithm
\citep{clo+09,brd+16,tch+18,zgz+19,tz20}.

We obtain the training sample of spectroscopic redshifts from the
HSC-SSP DR2, which includes the data from SDSS DR14 \citep{dr14+18},
DEEP3 \citep{cac+11}, PRIMUS DR1 \citep{cmb+13}, VIPERS PDR1
\citep{ggs+14}, VVDS \citep{lcc+13}, GAMA DR2 \citep{lbd+15}, WiggleZ
DR1 \citep{djb+10}, zCOSMOS DR3 \citep{llm+09}, UDSz
\citep{bah+13,mpd+13}, FMOS-COSMOS \citep{sks+15,kss+19} and 3DHST
\citep{swm+14,mbv+16}. To have enough data for training at $z>1$, we
also supplement the spectroscopic redshifts with accurate photometric
redshifts in the COSMOS2015 catalogue, which are based on 30-band
photometry with an accuracy of 0.021 \citep{lmi+16}. We adopt the
quality flags of the redshift data following \citet{tch+18}. This
spectroscopic sample contains 554,996 galaxies, of which 240,409 are
matched with the HSC-SSP$\times$unWISE galaxies and have magnitudes in
the $grizyW1$ bands. This obtained spectroscopic sample is biased to
low redshifts due to the data from the SDSS. To reduce the bias, we
randomly remove some of galaxies of $i<21.8$ \citep[the limit of SDSS
  redshift survey,][]{pln+16} to make that the training sample has
similar colour distributions with the photometric sample following
\citet{lco+08}. In the colour space ($g-r$, $r-i$, $i-z$, $z-y$ and
$i-{\rm W1}$), we calculate a probability,
  \begin{equation}
    p=\frac{N_{\rm S,tot}N_{\rm P}(\Delta c)}{N_{\rm P,tot}N_{\rm S}(\Delta c}.
  \end{equation}
Here, $N_{\rm S,tot}$ and $N_{\rm P,tot}$ are the total numbers of
galaxies in the spectroscopic and photometric samples, respectively.
$N_{\rm S}(\Delta c)$ and $N_{\rm P}(\Delta c)$ are the numbers of
neighbour galaxies within the colour interval $\Delta c=0.05$ (for all
five colours) of the galaxy from the spectroscopic and photometric
samples, respectively. If $p<1$, the galaxy has the probability of
$1-p$ being removed from the spectroscopic sample. We perform this with
the help of a random number generator. After randomly
removing the low-redshift galaxies, we adopt the remaining 208,164
galaxies as the training sample.
To evaluate the accuracy of photometric redshift, we randomly select
30,000 galaxies as a test sample and take the rest much larger sample
of 178,164 galaxies for training.

Galaxy colour is tightly related to redshift especially when the
survey bands cover the feature of 4000~\AA~break or 1.6 $\mu$m
bump. In the multidimensional colour space, the galaxies at the same
locations generally have similar redshifts. For the photometric
redshift of a target galaxy, we calculate the distance in the colour
space ($g-r$, $r-i$, $i-z$, $z-y$, $i-{\rm W1}$ and $i-{\rm W2}$) to
all the galaxies in the training sample. There are $k$ galaxies found
as the nearest neighbours in the colour space. The median value of
their spectroscopic redshifts is taken as the photometric redshift of
the target galaxy, and the scatter of the spectroscopic redshifts of
$k$ galaxies is taken as the error of estimated photometric
redshift. We test various number of $k$, and find that $k=20$ can lead
to the smallest scatter between the photometric redshifts and the
spectroscopic redshifts for the 30,000 testing galaxies. The
uncertainty of the photometric redshift in our work is defined as
$\sigma_{\Delta z}=1.48\times {\rm median}(|z_p-z_s|/(1+z_s))$. The
lower panel of Fig.~\ref{photoz} shows that the redshift uncertainty
is about 0.024 at $z<0.7$ and steadily increases with redshift to
about 0.11 at $z\sim2$ roughly following a formula of $\sigma_{\Delta
  z}=0.055z-0.0145$.  We consider the photometric redshifts with a
deviation larger than 3\,$\sigma_{\Delta z}$ or larger than 0.15 as
outliers, which is about 6\% for our photometric redshift estimates.

The HSC-SSP has already published photometric
redshifts\footnote{https://hsc-release.mtk.nao.ac.jp/doc/index.php/photometric-redshifts/}
for all HSC-SSP galaxies based on the five band ($grizy$) data. The
photometric redshift catalogue of DR1 \citep{tch+18} contains six
estimates based on one template fitting method \citep[Mizuki;][]{tan15}
and five neural network or machine learning methods, i.e. Direct
Empirical Photometric code \citep[DEmP;][]{hy14}, Extended Photometric
redshift, Flexible Regression over Associated Neighbours with
Kernel dEnsity estimatioN for Redshifts, MLZ \citep{cb14}
and Nearest Neighbours Photometric Redshift \citep{clo+09},
respectively. The photometric redshifts of the HSC-SSP DR2 galaxies
are provided within $z<5$ based on the DEmP and Mizuki methods
\citep{nht+20}. The redshift estimates of the HSC-SSP have an
uncertainty of about 0.05 and an outlier rate of about 15\% for the
galaxies down to $i=25$. In Fig.~\ref{photoz}, we compare both
photometric redshifts in the HSC-SSP DR2 (Mizuki estimate) and in this
work with spectroscopic redshifts. The HSC-SSP DR2 redshift has a
larger uncertainty at $1.4<z<2$ mainly because the 4000~\AA~break
moves out of the $y$ band, which can be improved with the help of the
unWISE data. 

We then take all 208,164 galaxies as the training sample, and obtain
photometric redshifts for 14.68 million HSC-SSP$\times$unWISE
galaxies\footnote{All data are available at
  http://zmtt.bao.ac.cn/galaxy\_clusters/} using the
nearest-neighbour algorithm.  The redshift distribution of the
HSC-SSP$\times$unWISE galaxies has a peak at $z\sim0.7$ and extends to
$z>2$, as shown in Fig.~\ref{histgz}.
The deeper COSMOS2015 data can be used to estimate the completeness of
the HSC-SSP$\times$unWISE data. The completeness is defined to be the
percentage calculated from the number of galaxies in
HSC-SSP$\times$unWISE catalogue over the number of galaxies in the
COSMOS2015 catalogue. As shown in Fig.~\ref{galcomplete}, the
completeness of galaxies depends on stellar mass and redshift. For the
galaxies with a stellar mass of $m_{\rm stellar}\sim
10^{11}~M_{\odot}$, the completeness decreases from $>90\%$ at $z<1$
to $\sim70\%$ at $z\sim1.5$. For the galaxies with a stellar mass of
$m_{\rm stellar}\sim 10^{10.5}~M_{\odot}$, the completeness drops from
$\sim90\%$ at $z<0.5$ to $\sim 70\%$ at $z\sim1$. The
  completeness may vary across the sky because the unWISE catalogue is
  not homogeneous and is relatively shallow in the COSMOS field
  \citep{smg19}.  The completeness of galaxies in other regions,
however, currently cannot be evaluated.

\begin{figure}
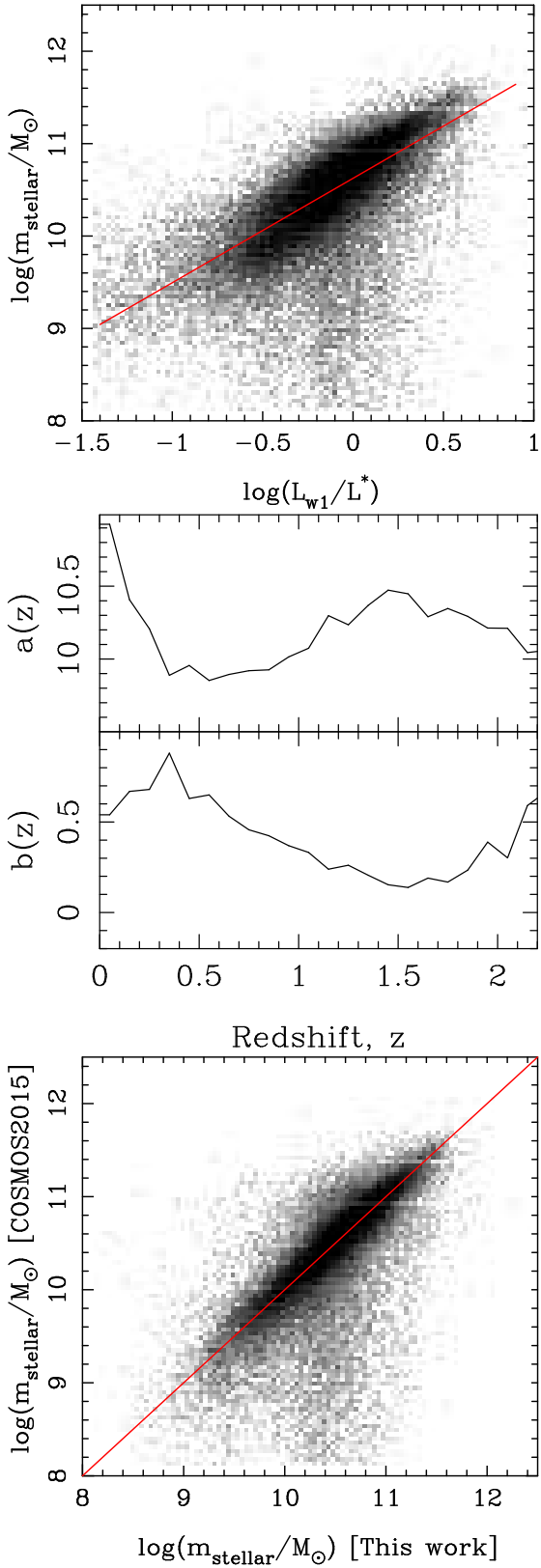

\centering
\includegraphics[width=0.41\textwidth]{f4a.eps} \hspace{3mm}
\includegraphics[width=0.41\textwidth]{f4b.eps} \hspace{3mm}
\includegraphics[width=0.41\textwidth]{f4c.eps}
\caption{Upper: correlation between galaxy stellar mass and ${\rm
    W1}$-band luminosity. Middle: The slope and intercept of the
  relation between stellar mass and galaxy colour as a function of
  redshift, which are used for improvement of the stellar
  mass--luminosity relation. Lower: comparison between the stellar
  mass in the COSMOS2015 catalogue and the estimate in this work.}
\label{galmass2}
\end{figure}

\begin{figure}
\centering \includegraphics[width = 0.45\textwidth]{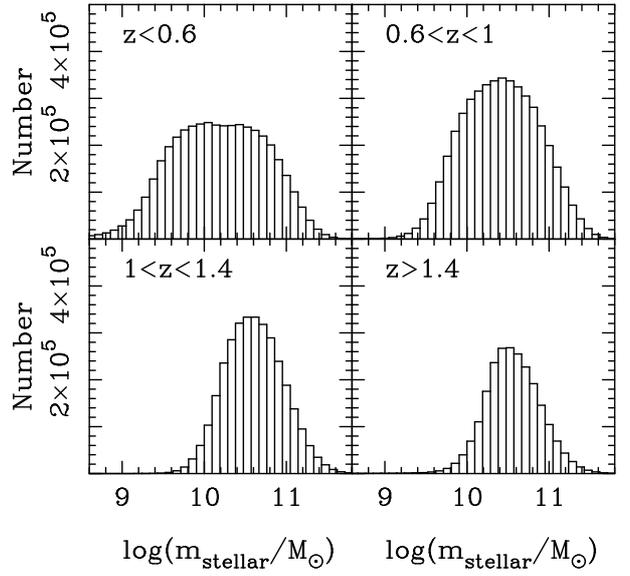}
\caption{Distribution of stellar masses of HSC-SSP$\times$unWISE galaxies
  at different redshift ranges.}
\label{histmstar}
\end{figure}

\subsection{Stellar masses of galaxies}
\label{galaxystellar}

The stellar masses ($m_{\rm stellar}$) of galaxies have a tight
correlation with infrared luminosities
\citep{bd01,dbh04,rhp+06,lwc+07}. We here derive the stellar mass
based on the ${\rm W1}$-band luminosity of galaxies. First we convert
the ${\rm W1}$-band magnitude to a luminosity by
\begin{equation}
  \log(L_{\rm W1}/L^{\ast})=-0.4({\rm W1}-{\rm W1}^{\ast}),
  \label{maglum}
\end{equation}
where ${\rm W1}^{\ast}$ is the reference characteristic magnitude from
a passive population synthesis model of \citet{bc03}, normalized to
${\rm W1}^{\ast}=16.45$ at $z=0.1$ \citep{mgb+10,ggb+19}. In the
evolution model, we assume that the stellar population was formed at a
redshift of $z_f=3$, and adopt the stellar evolution tracks of Padova
1994 \citep{gbc+96}, the `Basel3.1' stellar spectral library
\citep{wlb+02}, and the initial mass function of \citet{cha03} and the
Solar metallicity. The $L^{\ast}$ is the corresponding characteristic
luminosity.

To calibrate the relation between galaxy stellar mass and infrared
luminosity, we use the stellar mass data in the COSMOS2015 catalogue,
which are based on 30-band photometry for half million galaxies to a
limit of $K_s=24.7$ \citep{lmi+16}. Cross-matching the COSMOS2015
catalogue with the HSC-SSP$\times$unWISE catalogue gives 33,879 common
galaxies at $z<2.5$. We show the correlation between $L_{\rm
    W1}$ and $m_{\rm stellar}$ for the 33,879 galaxies (the upper
  panel of Fig.~\ref{galmass2}) and fit the relation to a power law
  for the whole sample,
\begin{equation}
  \log(m_{\rm stellar}/M_{\odot})=\gamma\log(L_{\rm W1}/L^{\ast})+f.
  \label{stelum}
\end{equation}
The best fitting gives the slope of $\gamma=1.13\pm0.01$ and the
intercept of $f=10.63\pm0.01$. The slope is consistent with that
derived by \citet{wwz+13}.  The scatter of stellar masses, $1.48\times
{\rm median}(\Delta m_{\rm stellar})$, is about 0.32 dex as estimated
by Eq.~(\ref{stelum}). Considering redshift dependence from possible
physical evolution of the luminosity--stellar relation or any bias of
the quoted reference ${\rm W1}^{\ast}$ \citep{wh18} and the colour
dependence \citep{bd01,wwz+13,cjh+14}, we divide the galaxy sample
into small redshift bins, and fit the $L_{\rm W1}$--$m_{\rm stellar}$
relation to the power law in Eq.~\ref{stelum} but with the intercept
$f$ in the form of
\begin{equation}
  f=a(z)+b(z)(r-Z).
  \label{intercept}
\end{equation}
The values of $a(z)$ and $b(z)$ are used to diminish the redshift
dependence, and are derived within each redshift bin (see middle panels
of Fig.~\ref{galmass2}). The $(r-Z)$ in Eq.~\ref{intercept} is the
colour index ($Z$ is for the filter band to avoid confusion with
redshift in this equation). After such corrections, the accuracy of
estimated stellar mass is improved from 0.32 dex to 0.21 dex (see the
lower panel of Fig.~\ref{galmass2}). We therefore calculate the
stellar masses for 14.68 million HSC-SSP$\times$unWISE galaxies. As
shown in Fig.~\ref{histmstar}, the HSC-SSP$\times$unWISE galaxies are
more massive at higher redshifts in general.

\begin{figure}
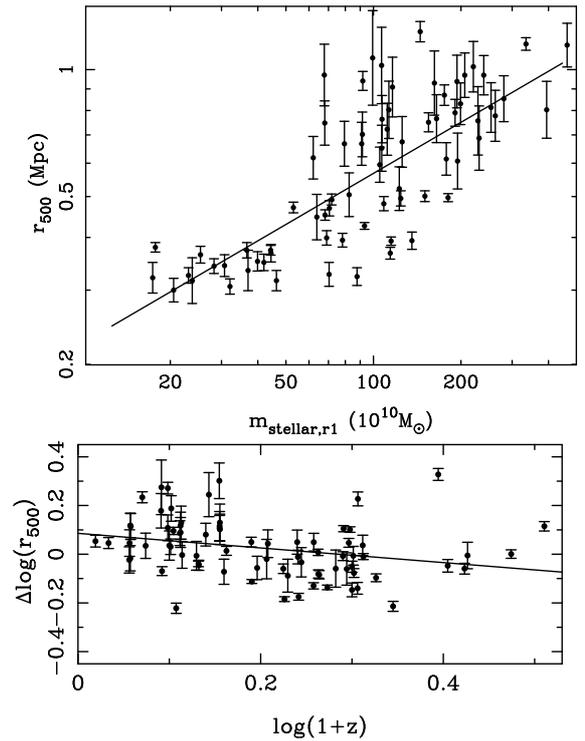

\centering
\includegraphics[width=0.41\textwidth]{f6a.eps} \hspace{3mm}
\includegraphics[width=0.41\textwidth]{f6b.eps} \hspace{3mm}
\caption{Upper: the scaling relation between cluster radius,
  $r_{500}$, and the total stellar mass, $m_{\rm stellar,r1}$, derived
  from HSC-SSP$\times$unWISE data for the test sample of 76 clusters from literature.
  The solid line is the best fit to the data. Lower: the deviation of $r_{500}$
  from the $r_{500}$--$m_{\rm stellar,r1}$ relation against $\log (1+z)$.}
\label{scaleXO}
\end{figure}

\begin{figure}
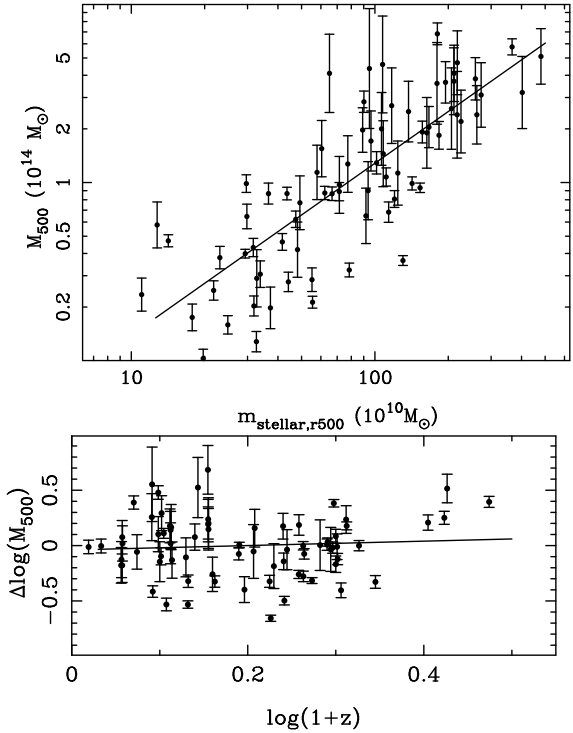

\centering
\includegraphics[width=0.41\textwidth]{f7a.eps} \hspace{3mm}
\includegraphics[width=0.41\textwidth]{f7b.eps} \hspace{3mm}
\caption{Similar to Fig.~\ref{scaleXO}, but for the
  $M_{500}$--$m_{\rm stellar,r500}$ relation determined for a test sample of clusters.}
\label{scaleX1}
\end{figure}

\section{Clusters of galaxies identified from the photometric redshift catalogue}

With the improved photometric redshifts of 14.68 million
HSC-SSP$\times$unWISE galaxies, we can identify clusters of galaxies
from the three-dimensional distribution of galaxies, following the
steps in our previous papers \citep{whl09,whl12,wh15}.  Here, we first
verify the scaling relations for the radius cluster and mass from
observables for a test sample of clusters, which will be used during
the identification of galaxy clusters.

\begin{table*}
\begin{minipage}{160mm}
\caption[]{The 21,661 clusters of galaxies identified from the
  HSC-SSP$\times$unWISE data.}
\begin{center}
\setlength{\tabcolsep}{1mm}
\begin{tabular}{crrrcccccrccc}
\hline
\mc{1}{c}{Cluster ID}& \mc{1}{c}{Name}&\mc{1}{c}{R.A.} & \mc{1}{c}{Dec.} & \mc{1}{c}{$z_{\rm cl}$} & \mc{1}{c}{$i_{\rm BCG}$} &
\mc{1}{c}{${\rm W1_{BCG}}$} &  \mc{1}{c}{SNR} & \mc{1}{c}{$r_{500}$} &
\mc{1}{c}{$\lambda_{500}$} & \mc{1}{c}{$M_{500}$} &\mc{1}{c}{$N_{\rm gal}$} & \mc{1}{c}{Other catalogues} \\
\mc{1}{c}{(1)} & \mc{1}{c}{(2)} & \mc{1}{c}{(3)} & \mc{1}{c}{(4)} & \mc{1}{c}{(5)} & 
\mc{1}{c}{(6)} & \mc{1}{c}{(7)} & \mc{1}{c}{(8)} & \mc{1}{c}{(9)} & \mc{1}{c}{(10)} &
\mc{1}{c}{(11)} & \mc{1}{c}{(12)} & \mc{1}{c}{(13)} \\
\hline
  1 & WH J000000.5$+$021911 & 0.00201 & $ 2.31980$ & 0.4297 & 18.791 & 17.964 &  6.73 & 0.560 & 21.16 & 0.96 & 14 &     \\
  2 & WH J000004.2$+$021941 & 0.01741 & $ 2.32800$ & 0.6165 & 19.557 & 17.721 &  6.02 & 0.656 & 30.51 & 1.36 & 17 & WHL \\
  3 & WH J000007.9$+$005153 & 0.03273 & $ 0.86473$ & 0.9170 & 21.493 & 18.930 &  5.17 & 0.491 & 20.54 & 0.93 & 10 &     \\
  4 & WH J000009.0$+$021817 & 0.03743 & $ 2.30465$ & 0.7390 & 19.643 & 17.888 &  6.32 & 0.610 & 28.34 & 1.27 & 15 &     \\
  5 & WH J000011.9$-$013226 & 0.04941 & $-1.54056$ & 1.3002 & 23.644 & 19.698 &  5.20 & 0.500 & 26.45 & 1.19 &  9 &     \\
  6 & WH J000014.7$+$005259 & 0.06120 & $ 0.88297$ & 0.6746 & 19.176 & 17.811 &  5.47 & 0.593 & 23.12 & 1.05 & 12 &     \\ 
  7 & WH J000015.0$-$003444 & 0.06253 & $-0.57879$ & 0.7652 & 20.286 & 17.729 &  7.53 & 0.610 & 32.70 & 1.46 & 12 &     \\ 
  8 & WH J000016.1$+$022955 & 0.06711 & $ 2.49857$ & 0.4365 & 18.671 & 17.910 &  5.53 & 0.625 & 17.41 & 0.80 & 13 & WHL \\ 
  9 & WH J000017.4$-$004841 & 0.07260 & $-0.81148$ & 0.6165 & 19.392 & 17.836 &  9.24 & 0.671 & 40.08 & 1.77 & 23 & WHL \\
 10 & WH J000020.3$+$014340 & 0.08467 & $ 1.72774$ & 0.8528 & 21.824 & 19.052 &  5.95 & 0.630 & 30.35 & 1.36 & 16 &     \\
\hline
\end{tabular}
\end{center}
{Note.
Column 1: Cluster ID;
Column 2: Cluster name with J2000 coordinates of cluster; 
Column 3 and 4: Right Ascension (R.A. J2000) and Declination (Dec. J2000) of cluster BCG (in degree);
Column 5: redshift of the cluster; 
Column 6--7: BCG magnitudes (AB system) in $i$ and ${\rm W1}$ bands, respectively;
Column 8: signal-to-noise ratio for cluster detection;
Column 9: cluster radius, $r_{500}$, in Mpc; 
Column 10: cluster richness.
Column 11: derived cluster mass, in units of $10^{14}~M_{\odot}$;
Column 12: number of member galaxy candidates within $r_{500}$; 
Column 13: Reference notes for 6047 previously known clusters: WH11 \citep{wh11}, WHL \citep{whl12,wh15},
CAMIRA14 \citep{ogu14}, CAMIRA18 \citep{oll+18}, WH18 \citep{wh18}, XXL \citep{agk+18},
ACT \citep{hhs+18}, MaDCoWS \citep{ggb+19}.\\
(This table is available in its entirety in a machine-readable form.)
}
\label{tab1}
\end{minipage}
\end{table*}

\subsection{The scaling relations for cluster radius and mass}
\label{scale}

Mass and radius are fundamental parameters of a cluster. The common
used radius, $r_{500}$, is the radius within which the mean density of
a cluster is 500 times of the critical density of the universe, and
$M_{500}$ is the cluster mass within $r_{500}$. Determining the scaling relations
  with observables of member galaxies is a prior step in our cluster
identification algorithm. As done in our previous work,
we used the total luminosity of member galaxies as the cluster mass
proxy for the SDSS clusters \citep{wh15}. Here, we have the stellar masses
of galaxies derived already. We re-calibrate the scaling relation between
the cluster mass (and radius) and the total stellar mass in clusters.

Following our previous papers \citep{whl12,wh15}, we use the clusters
with known $M_{500}$ (and $r_{500}$) as the test sample to get the
scaling relations. The HSC-SSP$\times$unWISE data cover the fields of
the deep X-ray surveys, e.g. the Chandra All-wavelength Extended Groth
Strip International Survey \citep{lng+09}, XMM-LSS survey
\citep{pva+04} and XMM XXL survey \citep{ppa+16}. We find the cluster
masses of $M_{500}$ for 76 known clusters in the HSC-SSP$\times$unWISE
data \citep{eft+13,pcg+16,agk+18,wrm+18}. The locations of BCGs are
taken as the centres of these clusters. We discriminate the member
galaxy candidates with a stellar mass of $m_{\rm stellar}\ge
5\times10^9~M_\odot$ in the HSC-SSP$\times$unWISE data if they are
within the redshift slice of $z\pm\Delta z$. Here, $\Delta z$ is
adopted in a simple form of
\begin{equation}
  \Delta z=\left\{
\begin{array}{ll}
   0.04\,(1+z)           &     \mbox{for $z\leq 0.7$}\\
   0.15\,z-0.037       &     \mbox{for $z> 0.7$}
\end{array}
\right. \,,
\label{pzslice}
\end{equation}
which is about 1.5\,$\sigma_{\Delta z}$ at $z<1.4$. Within this
photometric redshift slice, most member galaxies of a cluster can be
included with a relatively small contamination from field galaxies
that will be subtracted on average later \citep{whl09}.

To get the scaling relation for $r_{500}$, the total stellar mass is
calculated by summing the stellar masses of so-recognized member galaxy
candidates within a temporary radius, with a local background
subtracted. Instead of using a radius of 1~Mpc as in our previous papers
\citep{whl12,wh15}, we adopt the radius as
\begin{equation}
r_1=1.0\,E(z)^{-1/3}~{\rm Mpc},
\end{equation}
where $E(z)=\sqrt{\Omega_{\Lambda}+\Omega_m(1+z)^3}$, because the
radius of a $M_{500}$-fixed cluster changes with redshift by a factor
of $E(z)^{-1/3}$. The local background of the stellar mass is estimated
from an annulus of projected distance between 2 and 4 Mpc from the BCG
within the same redshift slice. We find that the $r_{500}$
  is well related to the total stellar mass $m_{\rm stellar,r1}$
  within $r_1$ following a power law (upper panel of
  Fig.~\ref{scaleXO}),
\begin{equation}
  \log r_{500}=\alpha\log m_{\rm stellar,r1}-\beta,
\label{r500stell}
\end{equation}
where $r_{500}$ is in units of Mpc and $m_{\rm stellar,r1}$ is in
units of $10^{10}$~M$_{\odot}$. The direct fitting by this small test
sample of X-ray flux-limited clusters usually suffers from the
Malmquist bias \citep{mae+10,oll+18}. We therefore prefer to fixing
the slope of this scaling relation according to the result of
\citet{wh15} obtained from a much large sample of clusters, which is
0.45 for the cluster radius--luminosity relation in the SDSS $r$
band. We assume that the slopes of the stellar mass--luminosity
relation in the $r$ and {\rm W1} bands are consistent
\citep{bsh+10}. Considering such an additional slope of 1.13
(Eq.~\ref{stelum}), we get the slope of $\alpha=0.45/1.13=0.40$ and
find the best fitting of $\beta=1.05\pm0.02$.
The deviations of $r_{500}$ from the radius--stellar mass relation,
$\Delta\log r_{500}=\log r_{500}-(0.40\log m_{\rm stellar,r1}-1.05)$,
have a negative redshift dependence in the form of (lower panel of
Fig.~\ref{scaleXO})
\begin{equation}
  \Delta\log r_{500}=-0.33\pm0.12\log (1+z)+(0.09\pm0.03).
  \label{dr500z}
\end{equation}
Combination of equations~(\ref{r500stell}) and (\ref{dr500z}) indicates that 
$r_{500}$ can be derived from the total stellar mass and redshift by
\begin{eqnarray}
\log r_{500}&=&0.40\log m_{\rm stellar,r1}-(0.96\pm0.03)\nonumber\\
&&+(0.33\pm0.12)\log (1+z). 
\label{r500z}
\end{eqnarray}

To get the scaling relation for $M_{500}$, we calculate the total
stellar mass, $m_{\rm stellar,500}$, within the radius of $r_{500}$,
and find a reasonable correlation between $M_{500}$ (in units of
$10^{14}~M_{\odot}$) and $m_{\rm stellar,500}$ (upper panel of
Fig.~\ref{scaleX1}). \citet{wh15} got an accurate slope of 1.08 for
the cluster mass--luminosity relation in the $r$ band. We assume a
power law for the cluster mass--stellar mass relation and fix the
slope to be $1.08/1.13=0.96$ after considering the conversion with an
additional slope between luminosity and stellar mass. The best fit we
obtain is
\begin{equation}
  \log M_{500}=0.96\log m_{\rm stellar,500}-(1.82\pm0.02).
  \label{m500stell}
\end{equation}
The deviations of $M_{500}$ from the cluster
mass--stellar mass relation, $\Delta\log M_{500}=\log
M_{500}-(0.96\log m_{\rm stellar,500}-1.82)$, have a very weak
dependence on redshift in the form of (lower panel of
Fig.~\ref{scaleX1})
\begin{equation}
  \Delta\log M_{500}=(0.20\pm0.31)\log (1+z)-(0.04\pm0.07).
  \label{dm500z} 
\end{equation}
Combination of equations~(\ref{m500stell}) and (\ref{dm500z}) indicates that
$M_{500}$ can be derived from the total stellar mass and redshift by
\begin{eqnarray}
\log M_{500} & = & 0.96\log m_{\rm stellar,500}-(1.86\pm0.07) \nonumber \\   
            &   & + (0.20\pm0.31)\log (1+z).
\label{m500z}
\end{eqnarray}
Furthermore, we define the richness $\lambda_{500}$ as a mass proxy by
\begin{equation}
  \lambda_{500}=m_{\rm stellar,500}(1+z)^{0.21}/m^{\ast}_{\rm stellar},
\label{richdef}
\end{equation}
where $m^{\ast}_{\rm stellar}\sim 4\times10^{10}$~$M_{\odot}$ is the
mean stellar mass of a galaxy with the luminosity of $L^{\ast}$ at
$z\sim 0.1$. The cluster mass, $M_{500}$, is related to the richness,
$\lambda_{500}$, by
\begin{equation}
\log M_{500}=0.96\log \lambda_{500}-(1.29\pm0.02),
\label{m500rich}
\end{equation}
which is redshift independent and generally in agreement with that in
our previous work \citep{wh15}.

\begin{figure}
\centering
\includegraphics[width = 0.4\textwidth]{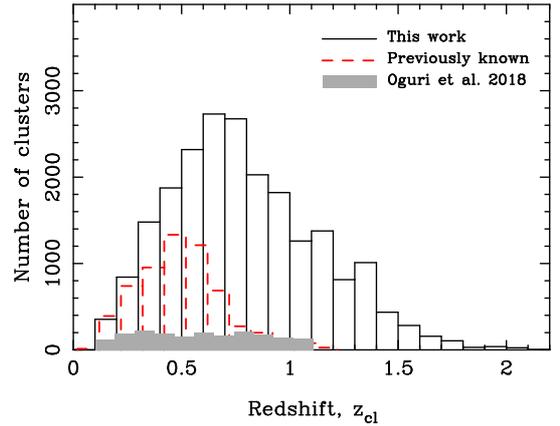}
\caption{Redshift distribution of the 21,661 clusters identified from
  the HSC-SSP$\times$unWISE data. The dashed histogram indicates the
  distribution of 6047 clusters previously known. The grey is for the
  CAMIRA clusters from HSC Wide S16A data release \citep{oll+18}.}
\label{hist_zc}
\end{figure}

\begin{figure*}
\resizebox{58mm}{!}{\includegraphics{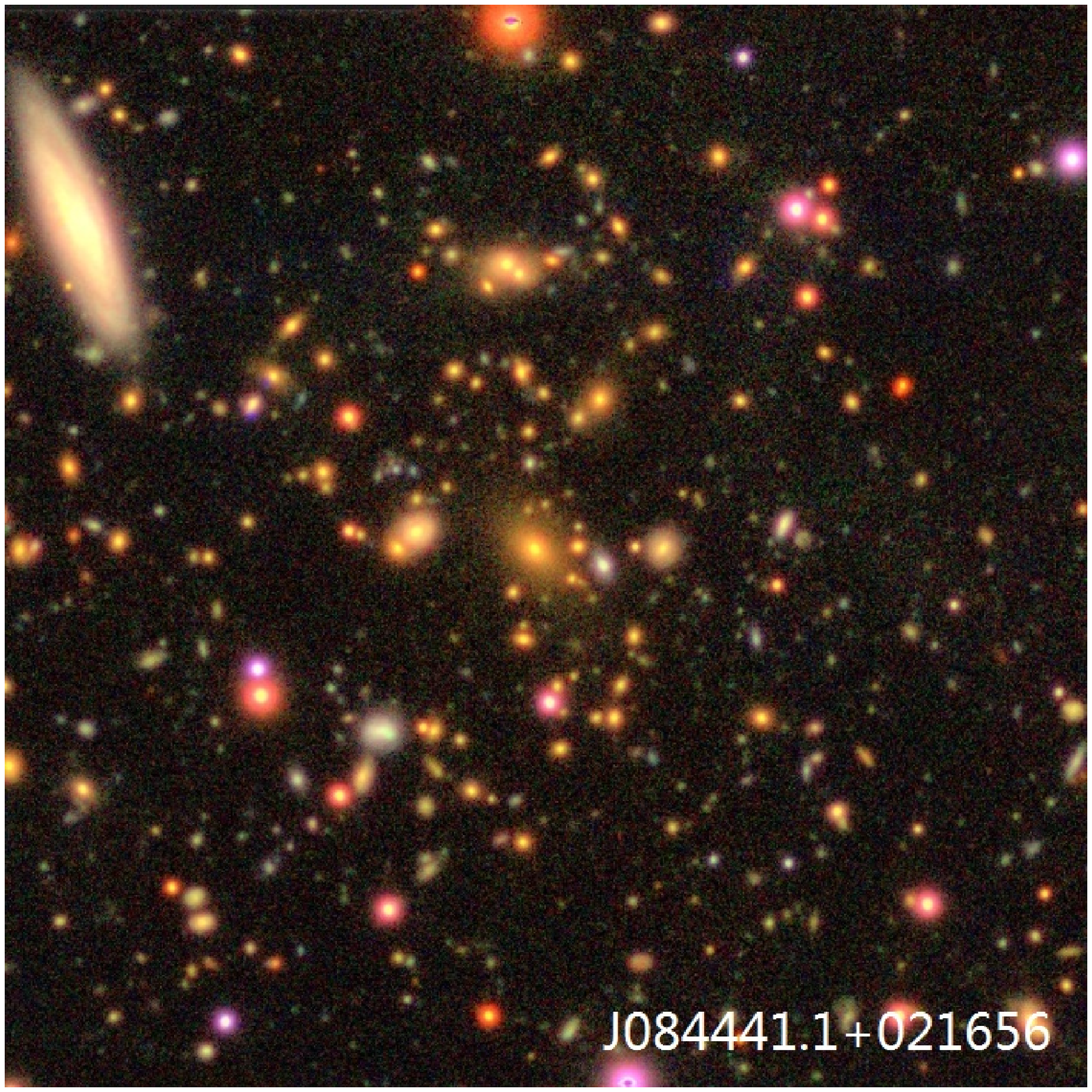}}
\resizebox{58mm}{!}{\includegraphics{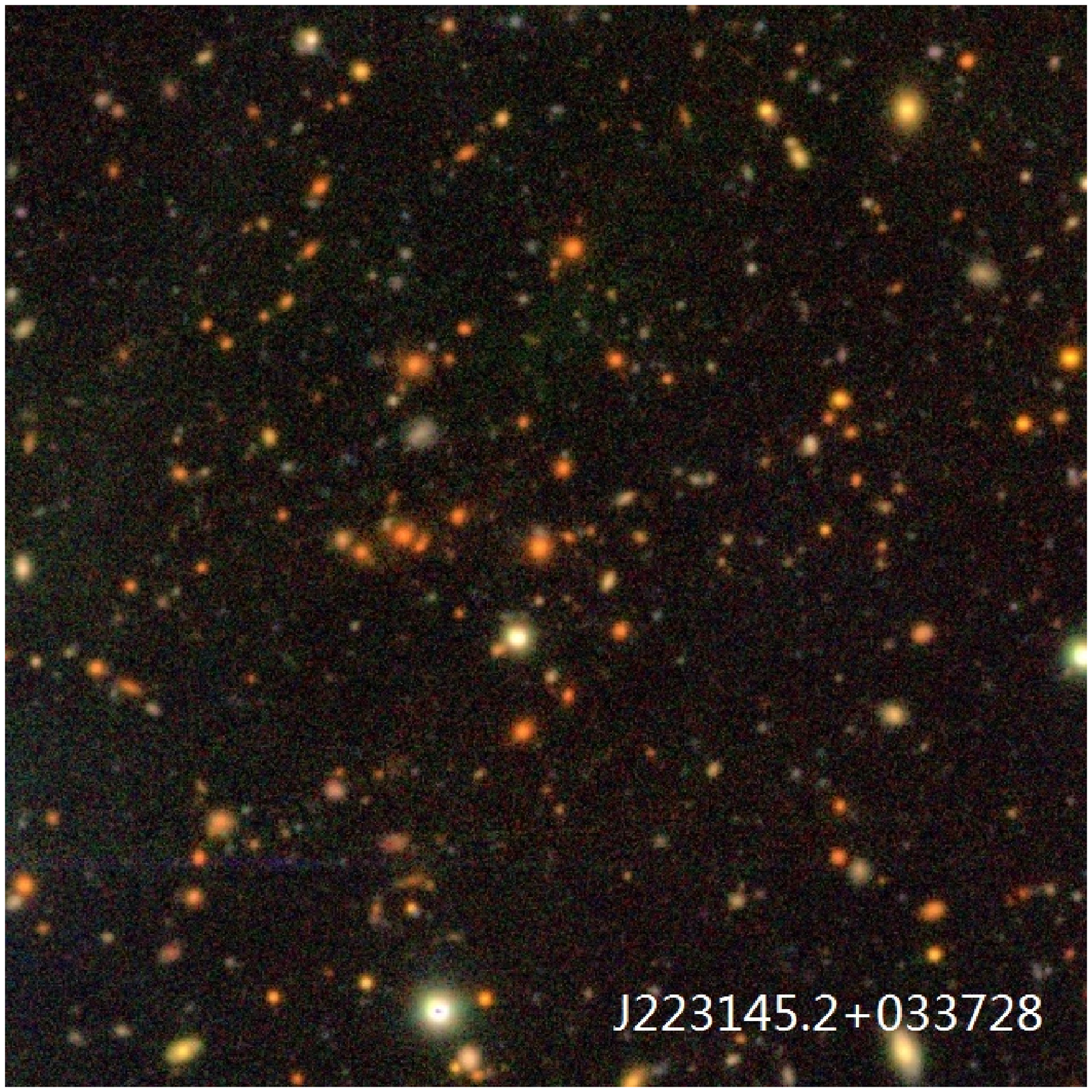}}
\resizebox{58mm}{!}{\includegraphics{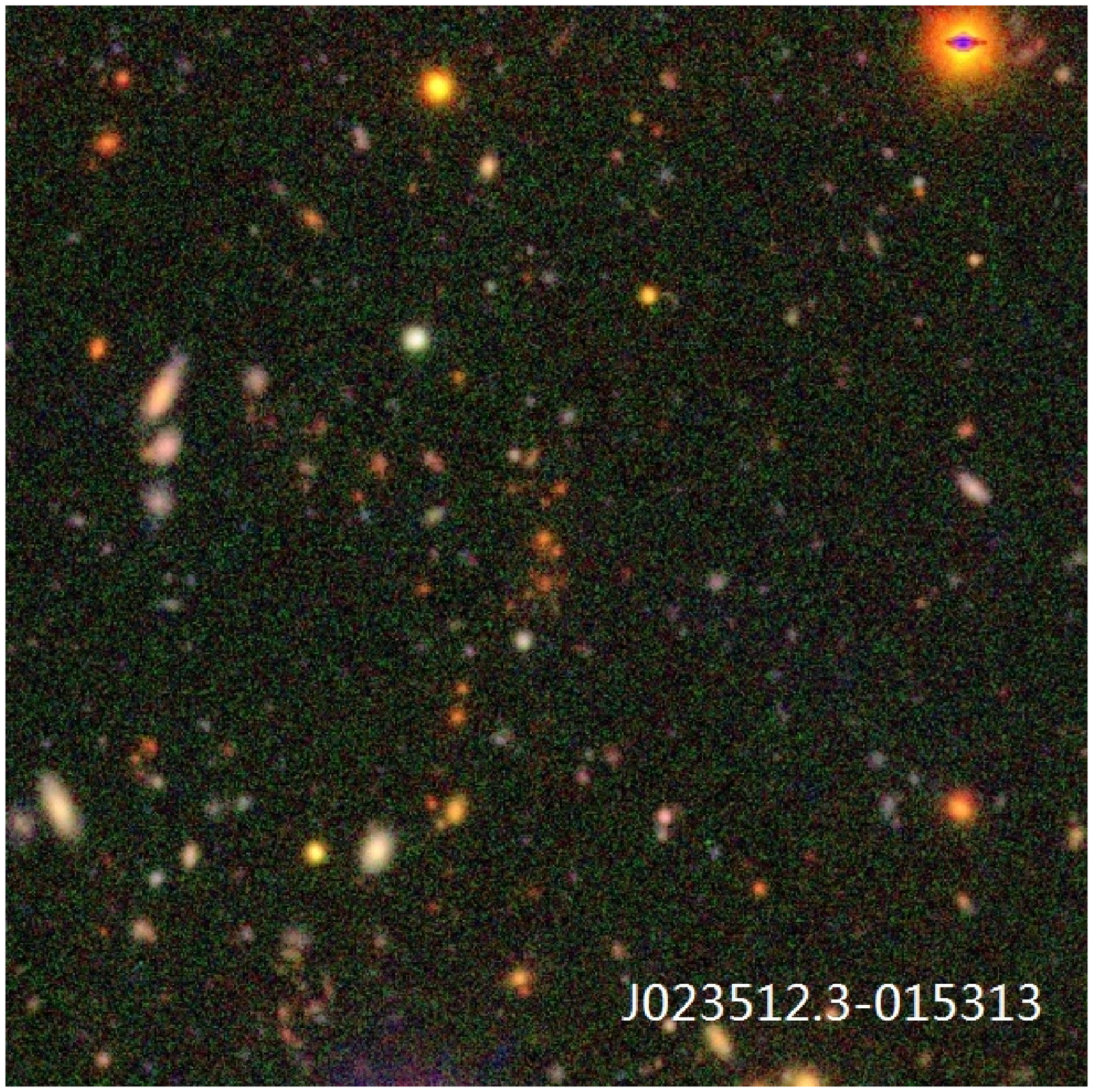}}
\caption{HSC-SSP composite ($riz$) colour images of three clusters at
  $z_{\rm cl}=0.6579$ (left), 0.9986 (middle) and 1.4594 (right), respectively,
  identified from the HSC-SSP$\times$unWISE data. The images have a
  scale of 1\,Mpc$\times$1\,Mpc.}
\label{example}
\end{figure*}

\begin{figure}
  \centering \includegraphics[width = 0.4\textwidth]{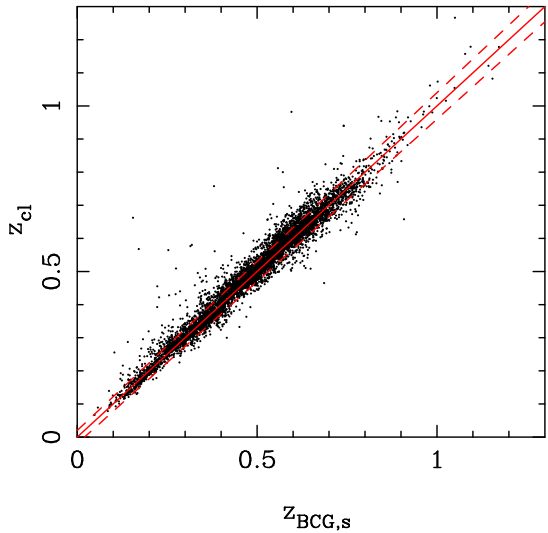}
  \caption{Comparison between the estimated cluster redshifts with
    spectroscopic redshifts of 6310 BCGs. The red dashed lines
    indicate the accuracy as being $0.022(1+z)$.}
\label{spzc}
\end{figure}

\subsection{Identification of galaxy clusters}
\label{algorithm}

In optical/infrared data, galaxy clusters are indicated by a high
overdensity of member galaxies. Since galaxy population in
  clusters is dominated by massive galaxies with a high stellar mass,
  the total stellar mass of member galaxies is not seriously affected
  by the incompleteness of low-mass galaxies. Using the estimated
photometric redshifts and stellar masses of galaxies, we can identify
galaxy clusters from the HSC-SSP$\times$unWISE galaxies following the
approach in our previous papers \citep{whl09,whl12,why18}, which
includes following four steps:

1. Estimate redshifts for cluster candidates.  Clusters usually
  contain BCGs with a high stellar mass \citep{lhl+17}. We
temporarily assume that each massive galaxy of $m_{\rm
  stellar}\ge10^{11}~M_{\odot}$ is the BCG of a cluster candidate at
$z$. The member galaxy candidates with a stellar mass of $m_{\rm
  stellar}\ge 5\times10^9~M_\odot$ are searched from the galaxies
within the photometric redshift slice of $z\pm\Delta z$ and the radius
of $r_1$. The cluster redshift is taken to be the median value of the
photometric redshifts of these member galaxy candidates.

2. Calculate a signal-to-noise ratio of the overdensity for cluster
candidates. For each cluster candidate, we get the sum of stellar mass
of member galaxy candidates within a projected radius of 0.5 Mpc from
the BCG, $m_{\rm stellar,0.5}$. We also estimate a local
``background'', $\langle m_{\rm stellar,0.5}\rangle$, and a
``fluctuation'', $\sigma_{m_{\rm stellar,0.5}}$, of the stellar mass
within the same redshift slice \citep{why18,wh18}. The
``signal-to-noise ratio'' of an overdensity region is defined as ${\rm
  SNR} = (m_{\rm stellar,0.5}-\langle m_{\rm
  stellar,0.5}\rangle)/\sigma_{m_{\rm stellar,0.5}}$. A larger $\rm
SNR$ means more stellar mass of galaxies concentrated on the BCG
candidates, which indicates a higher likelihood of true clusters.

3. Derive the radius and richness of cluster candidates based on the
scaling relations obtained in Section~\ref{scale}. We first calculate
the total stellar mass, $m_{\rm stellar,r1}$, within the radius of
$r_1$ from the HSC-SSP$\times$unWISE data, so that $r_{500}$ is
derived using the Eq.~(\ref{r500z}). Then, we calculate the total
stellar mass, $m_{\rm stellar,500}$, within the estimated radius
$r_{500}$.  The cluster richness, $\lambda_{500}$, is derived using
the Eq.~(\ref{richdef}).

4. Clean the entries. A rich cluster usually contains several massive
galaxies with $m_{\rm stellar}\ge10^{11}~M_{\odot}$, so that a cluster
can be repeatedly identified in the above procedures. The repeated
entries have to be merged into one cluster if they have a redshift
difference smaller than 1.5\,$\Delta z$ and a projected distance
smaller than $1.5\,r_{500}$. The entry with the largest richness is
then adopted for such a multi-identified cluster.

After these steps, we select only the clusters with a high overdensity
of ${\rm SNR}\ge5$ and a richness of $\lambda_{500}\ge15$. The masses
of so identified clusters have $M_{500}\ge 0.7\times
10^{14}~M_{\odot}$ according to the Eq.~(\ref{m500rich}). To avoid false
detections with very few member galaxies at high redshifts, we set the
threshold for the number of member galaxy candidates within $r_{500}$
as being $N_{\rm gal}\ge6$. The clusters of $z_{\rm cl}<0.1$ are also
discarded because bright HSC-SSP galaxies are possibly saturated and
are not included in the HSC-SSP$\times$unWISE data. Finally, we get
21,661 clusters from the HSC-SSP$\times$unWISE data, as listed in
Table~\ref{tab1}, among which 6047 clusters are previously known
\citep{wh11,whl12,wh15,ogu14,oll+18,wh18,agk+18, hhs+18,ggb+19}.
Fig.~\ref{hist_zc} shows that the identified clusters have a redshift
in the range of $0.1<z_{\rm cl}\lesssim2$ with a peak at $z_{\rm
  cl}\sim0.7$. There are 5537 clusters at $z_{\rm cl}\ge1$ and 642
clusters at $z_{\rm cl}\ge1.5$. About 90\% of clusters are newly
identified at $z_{\rm cl}>0.7$. In Fig.~\ref{example}, we show three
identified clusters from the HSC-SSP$\times$unWISE data at $z_{\rm
  cl}=0.6579$, 0.9986 and 1.4594, respectively. The cluster at $z_{\rm
  cl}=0.6579$ shows many member galaxies and has the BCG with
magnitudes of $i=19.260$ and ${\rm W1}=17.659$. The cluster at $z_{\rm
  cl}=1.4594$ has the BCG with magnitudes of $i=22.757$ and ${\rm
  W1}=18.922$. Considering the data limit of the unWISE, we can see
the concentration of the member galaxies 1.5 mag fainter than the BCG
in this cluster centre. By comparing with available spectroscopic
redshifts of 6310 BCGs, we find that the estimated cluster redshift
has an accuracy of $(z_{\rm cl}-z_{\rm BCG,s})/(1+z_{\rm BCG,s})$, which is
about $0.022$ (see Fig.~\ref{spzc}). The member candidates of 372,974
galaxies within $r_{500}$ for the 21,661 clusters are listed in
Table~\ref{tab2}.

\begin{landscape}
\begin{table}
%\scriptsize
\caption[]{The member candidates of 372,974 galaxies for 21,661 identified clusters from the
  HSC-SSP$\times$unWISE data.}
\begin{center}
\begin{tabular}{cccccccccccccccccrcc}
\hline
\mc{1}{c}{CluID}&\mc{1}{c}{R.A.} & \mc{1}{c}{Dec.} & \mc{1}{c}{$z$} & \mc{1}{c}{$g$} & \mc{1}{c}{${\rm err}_g$} &
\mc{1}{c}{$r$} & \mc{1}{c}{${\rm err}_r$} & \mc{1}{c}{$i$} & \mc{1}{c}{${\rm err}_i$} &
\mc{1}{c}{$Z$} & \mc{1}{c}{${\rm err}_Z$} & \mc{1}{c}{$y$} & \mc{1}{c}{${\rm err}_y$} &
\mc{1}{c}{${\rm W1}$} & \mc{1}{c}{${\rm err}_{\rm W1}$} & \mc{1}{c}{${\rm W2}$} & \mc{1}{c}{${\rm err}_{\rm W2}$} &
\mc{1}{c}{$\log(m_{\rm stellar})$} & \mc{1}{c}{$r_p$} \\
\mc{1}{c}{(1)} & \mc{1}{c}{(2)} & \mc{1}{c}{(3)} & \mc{1}{c}{(4)} & \mc{1}{c}{(5)} & \mc{1}{c}{(6)} &
\mc{1}{c}{(7)} & \mc{1}{c}{(8)} & \mc{1}{c}{(9)} & \mc{1}{c}{(10)} & \mc{1}{c}{(11)} & \mc{1}{c}{(12)} &
\mc{1}{c}{(13)} & \mc{1}{c}{(14)} & \mc{1}{c}{(15)} & \mc{1}{c}{(16)} & \mc{1}{c}{(17)} & \mc{1}{c}{(18)} &
\mc{1}{c}{(19)} & \mc{1}{c}{(20)} \\
\hline
     1 &   0.00009 &   2.30655 & 0.4775& 22.898 & 0.017& 21.751 & 0.006& 21.151 & 0.004& 20.828 & 0.007& 20.584 & 0.012& 19.885 & 0.060& 20.299 & 0.183 & 10.163& 0.270 \\
     1 &   0.00956 &   2.30905 & 0.4423& 22.835 & 0.016& 21.252 & 0.005& 20.508 & 0.003& 20.191 & 0.004& 19.997 & 0.008& 19.630 & 0.048& 99.000 &99.000 & 10.398& 0.265 \\
     1 &   0.01818 &   2.32395 & 0.4118& 22.593 & 0.011& 20.854 & 0.003& 20.073 & 0.002& 19.727 & 0.003& 19.546 & 0.004& 19.116 & 0.031& 19.564 & 0.098 & 10.732& 0.337 \\
     1 &   0.01162 &   2.29570 & 0.4373& 23.834 & 0.032& 22.331 & 0.009& 21.638 & 0.007& 21.394 & 0.013& 21.178 & 0.024& 20.819 & 0.137& 99.000 &99.000 &  9.793& 0.523 \\
     1 &   0.01247 &   2.30269 & 0.4400& 21.936 & 0.010& 20.238 & 0.003& 19.431 & 0.001& 19.061 & 0.002& 18.861 & 0.003& 18.442 & 0.018& 19.055 & 0.063 & 11.011& 0.405 \\
     1 &   0.02738 &   2.31081 & 0.3919& 21.991 & 0.007& 20.414 & 0.002& 19.724 & 0.001& 19.416 & 0.002& 19.250 & 0.004& 18.864 & 0.025& 19.457 & 0.089 & 10.784& 0.543 \\
     1 &   0.01404 &   2.31318 & 0.4292& 22.163 & 0.010& 20.452 & 0.003& 19.692 & 0.001& 19.339 & 0.002& 19.148 & 0.004& 18.036 & 0.013& 18.466 & 0.038 & 11.172& 0.277 \\
     1 &   0.00201 &   2.31980 & 0.4385& 21.167 & 0.007& 19.597 & 0.002& 18.791 & 0.001& 18.372 & 0.002& 18.121 & 0.003& 17.964 & 0.013& 18.254 & 0.032 & 11.259& 0.000 \\
     1 &   0.02213 &   2.32323 & 0.4832& 21.897 & 0.009& 20.787 & 0.004& 20.188 & 0.002& 19.837 & 0.004& 19.535 & 0.006& 18.743 & 0.023& 18.846 & 0.052 & 10.687& 0.412 \\
     1 &   0.02478 &   2.32432 & 0.4857& 22.286 & 0.010& 20.856 & 0.004& 20.087 & 0.002& 19.791 & 0.003& 19.533 & 0.006& 18.801 & 0.024& 19.049 & 0.063 & 10.732& 0.468 \\
     1 &   0.01072 &   2.34266 & 0.4389& 23.039 & 0.020& 21.482 & 0.005& 20.728 & 0.003& 20.403 & 0.005& 20.235 & 0.012& 20.035 & 0.068& 99.000 &99.000 & 10.235& 0.494 \\
     1 &   0.01372 &   2.32647 & 0.3770& 23.386 & 0.030& 22.030 & 0.010& 21.497 & 0.007& 21.209 & 0.012& 20.992 & 0.023& 20.625 & 0.113& 99.000 &99.000 &  9.876& 0.272 \\
     1 &   0.02198 &   2.32837 & 0.4442& 22.571 & 0.013& 20.863 & 0.003& 20.092 & 0.002& 19.746 & 0.003& 19.559 & 0.005& 18.605 & 0.020& 19.206 & 0.071 & 10.889& 0.438 \\
     1 &   0.00964 &   2.32917 & 0.3948& 22.652 & 0.012& 21.104 & 0.004& 20.421 & 0.002& 20.037 & 0.003& 19.853 & 0.006& 19.033 & 0.029& 19.263 & 0.076 & 10.757& 0.244 \\
     2 &   0.01831 &   2.30579 & 0.6107& 23.615 & 0.024& 21.711 & 0.006& 20.502 & 0.002& 20.078 & 0.003& 19.836 & 0.006& 18.814 & 0.024& 19.432 & 0.087 & 11.109& 0.541 \\
     2 &   0.02242 &   2.30766 & 0.6322& 23.368 & 0.036& 22.213 & 0.019& 21.889 & 0.017& 21.337 & 0.022& 21.687 & 0.041& 19.352 & 0.038& 19.593 & 0.100 & 10.419& 0.510 \\
     2 &   0.01087 &   2.31278 & 0.6132& 22.804 & 0.016& 21.673 & 0.007& 20.951 & 0.004& 20.723 & 0.007& 20.526 & 0.014& 19.656 & 0.050& 99.000 &99.000 & 10.335& 0.403 \\
     2 &   0.01134 &   2.31443 & 0.6628& 24.218 & 0.036& 22.516 & 0.009& 21.384 & 0.003& 20.955 & 0.006& 20.760 & 0.010& 19.568 & 0.046& 99.000 &99.000 & 10.666& 0.362 \\
     2 &   0.01732 &   2.31986 & 0.6291& 22.743 & 0.017& 22.316 & 0.011& 21.963 & 0.008& 21.978 & 0.020& 21.723 & 0.033& 20.073 & 0.070& 99.000 &99.000 &  9.798& 0.198 \\
     2 &   0.01042 &   2.32136 & 0.6165& 22.443 & 0.011& 21.643 & 0.006& 21.117 & 0.004& 20.935 & 0.008& 20.743 & 0.015& 20.308 & 0.086& 99.000 &99.000 &  9.903& 0.235 \\
     2 &   0.02610 &   2.31993 & 0.5636& 22.503 & 0.013& 21.005 & 0.004& 20.051 & 0.002& 19.674 & 0.003& 19.529 & 0.006& 18.618 & 0.021& 99.000 &99.000 & 11.073& 0.289 \\
     2 &   0.01256 &   2.33595 & 0.6015& 24.670 & 0.091& 23.380 & 0.030& 22.603 & 0.011& 22.368 & 0.021& 22.235 & 0.053& 21.077 & 0.168& 99.000 &99.000 &  9.743& 0.227 \\
     2 &   0.01741 &   2.32800 & 0.6630& 22.303 & 0.012& 20.770 & 0.004& 19.557 & 0.001& 19.087 & 0.002& 18.756 & 0.003& 17.721 & 0.011& 18.116 & 0.029 & 11.558& 0.000 \\
     2 &   0.01587 &   2.32853 & 0.6682& 23.655 & 0.026& 22.065 & 0.007& 20.885 & 0.003& 20.461 & 0.004& 20.141 & 0.007& 18.690 & 0.023& 19.011 & 0.062 & 11.078& 0.040 \\
     2 &   0.01787 &   2.32850 & 0.6675& 24.070 & 0.049& 22.193 & 0.010& 21.160 & 0.004& 20.737 & 0.007& 20.488 & 0.012& 18.652 & 0.024& 99.000 &99.000 & 11.018& 0.017 \\
     2 &   0.00335 &   2.33072 & 0.6702& 22.734 & 0.010& 21.801 & 0.005& 20.484 & 0.002& 20.120 & 0.003& 20.037 & 0.005& 19.430 & 0.041& 20.078 & 0.154 & 10.784& 0.348 \\
     2 &   0.02441 &   2.33236 & 0.5731& 23.923 & 0.034& 22.311 & 0.013& 21.265 & 0.004& 20.806 & 0.006& 20.550 & 0.012& 19.534 & 0.045& 19.719 & 0.112 & 10.760& 0.201 \\
     2 &   0.02096 &   2.33169 & 0.6769& 22.583 & 0.013& 21.593 & 0.006& 20.733 & 0.003& 20.387 & 0.005& 20.241 & 0.009& 19.078 & 0.030& 19.444 & 0.088 & 10.693& 0.125 \\
     2 &   0.04241 &   2.33564 & 0.6069& 22.696 & 0.035& 21.569 & 0.010& 20.604 & 0.004& 20.345 & 0.006& 20.150 & 0.011& 19.565 & 0.045& 20.042 & 0.147 & 10.540& 0.636 \\
     2 &   0.02966 &   2.33760 & 0.6790& 22.542 & 0.016& 21.765 & 0.007& 21.078 & 0.004& 20.832 & 0.008& 20.614 & 0.014& 19.542 & 0.045& 19.978 & 0.140 & 10.344& 0.379 \\
     2 &   0.02303 &   2.34174 & 0.6420& 25.040 & 0.084& 23.428 & 0.023& 22.320 & 0.008& 21.894 & 0.015& 21.642 & 0.025& 20.750 & 0.127& 99.000 &99.000 & 10.143& 0.361 \\
\hline                                                                                                                               
\end{tabular}
\end{center}
{Note. 
Column 1: Cluster ID in Table~\ref{tab1};
Column 2 and 3: Right Ascension (R.A. J2000) and Declination (Dec. J2000) of member galaxy candidate (in degree);
Column 4: photometric redshift of the galaxy;
Column 5--18: magnitude and error in the $g$, $r$, $i$, $Z$, $y$, ${\rm W1}$ and ${\rm W2}$ bands. The value of 99.000 means no data;
Column 19: logarithm of galaxy stellar mass, $\log(m_{\rm stellar}/M_{\odot})$;
Column 20: projected distance to cluster centre, in Mpc.
\\
(This table is available in its entirety in a machine-readable form.)
}
\label{tab2}
\end{table}
\end{landscape}

\begin{figure}
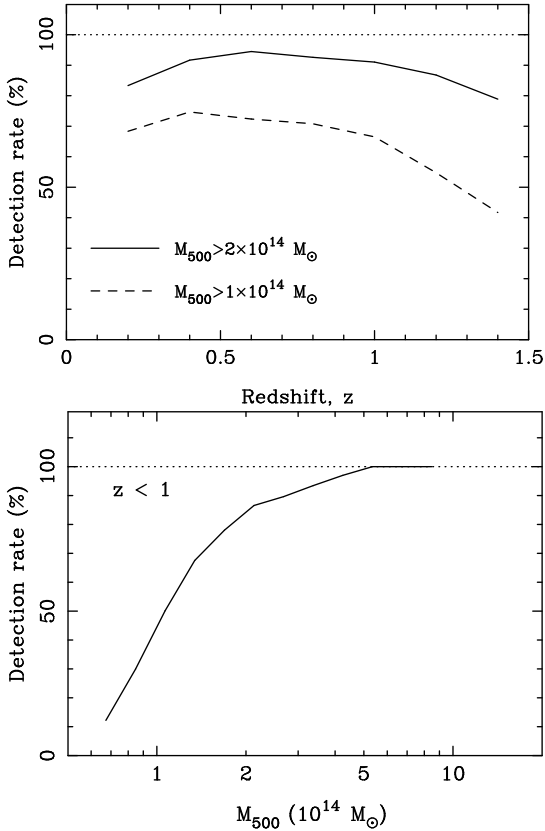

  \centering \includegraphics[width = 0.4\textwidth]{f11a.eps}
  \centering \includegraphics[width = 0.4\textwidth]{f11b.eps}
  \caption{Detection rate of mock clusters as a function
    of redshift (upper) and cluster mass (lower) for completeness tests.}
\label{clucomplete}
\end{figure}

\subsection{Completeness estimate}
\label{completeness}

Completeness is one of necessary evaluations for an optical cluster
catalogue. It can be estimated by two methods: the
cross-matching with X-ray cluster catalogues and the Monte-Carlo
simulations. The former is usually used for low redshift clusters
\citep[e.g.][]{ogu14,why18}. The latter uses mock clusters based on
simulations \citep[e.g.][]{kma+07b,whl09,hmk+10,whl12,rrb+14,oll+18}.
Here, Monte-Carlo simulations are performed to estimate completeness
of our cluster catalogue following our previous work \citep{whl12}.

In the HSC-SSP field, we generate a population of mock clusters with a
mass function of \citet{tkk+08} within a volume of $z<1.5$. The total
stellar mass inside a cluster is derived following the cluster
mass-richness relation
\begin{equation}
  \log M_{500}=0.96\log (C\times \lambda_{500})-1.29,
\label{m500in}
\end{equation}
where the factor $C$ is the ratio between the input and output
richness due to the uncertainty of galaxy photometric redshift. The
value of $C$ is to be determined so that the output richness follows
the Eq.~(\ref{m500rich}). For each mock cluster, the member galaxy
population is obtained following the stellar mass function by
\citet{vrm+18}. The photometric redshift of each member galaxy is given as
the cluster redshift ($z$) plus an error, i.e. $z+z_{\rm err}$, where $z_{\rm err}$
is a Gaussian random number with a scatter of the redshift uncertainty
shown in Fig.~\ref{photoz}. The surface profile of member galaxies
follows a two-dimensional NFW profile \citep{bar96} with a
concentration of $c_{500}=2$ \citep{wh18}. The final mock member
galaxies are obtained by considering the completeness of galaxies as
functions of redshift and stellar mass shown in
Fig.~\ref{galcomplete}.

The completeness simulation is performed by two steps. First, we set
$C=1$ to generate the mock clusters and their member galaxies. After
adding the mock member galaxies into the observed
HSC-SSP$\times$unWISE data, we apply the procedure of cluster
identification in Section~\ref{algorithm} to identify mock
clusters. The output richness is scaled to the input richness by
$C=1/0.82$ \citep{whl12}. Secondly, we re-generate the mock member
galaxies using the Eq.~(\ref{m500in}) with the new value of $C$ and
add them into the HSC-SSP$\times$unWISE data. Again, the procedure of
cluster identification in Section~\ref{algorithm} is applied to
identify the mock clusters.
The detection rate of mock clusters is regarded as a completeness
estimate of our cluster catalogue, which is defined to be the ratio
between the number of identified mock clusters and the total number of
mock clusters. We show in the upper panel of Fig.~\ref{clucomplete}
that the completeness slightly varies with redshift. For massive
clusters of $M_{500}>2\times10^{14}~M_{\odot}$, the completeness is
$\gtrsim90\%$ up to $z\sim 1$ but decreases to $\sim80\%$ at $z\sim
1.5$. The completeness is about 70\% up to $z\sim 1$ for clusters of
$M_{500}>1\times10^{14}~M_{\odot}$. Figure~\ref{clucomplete} also
shows the completeness increases with cluster mass at $z<1$,
from 50\% at $M_{500}\sim 1\times10^{14}~M_{\odot}$ to 85\% at
$M_{500}\sim 2\times10^{14}~M_{\odot}$ (lower panel).

\begin{figure}
  \centering
  \includegraphics[width = 0.4\textwidth]{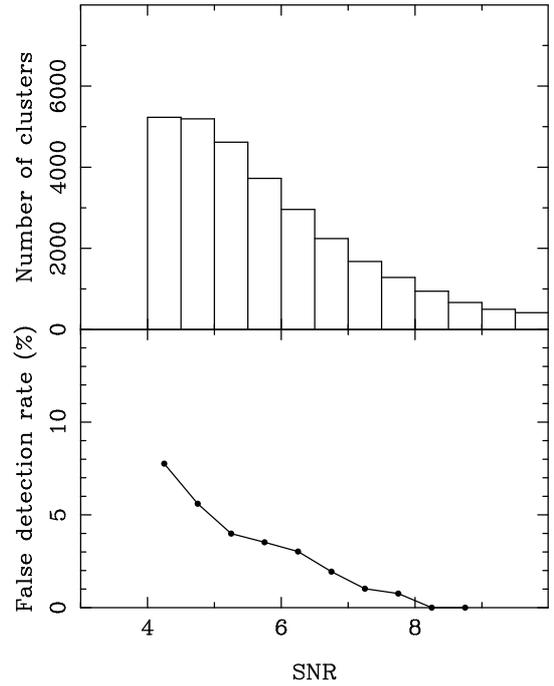}
\caption{Distribution of cluster detection ${\rm SNR}$ (upper)
  and the false detection rate against ${\rm SNR}$ (lower).}
\label{falserate}
\end{figure}

\subsection{False detection rate}
\label{false}

The filament structures along the line of sight show some overdensity
regions of galaxy stellar mass. Due to the large photometric redshift
slice we adopt, the projection effect may result in false detection
of galaxy clusters. 

Following previous works \citep{gsn+02,hmk+10,whl12}, we estimate the
false detection rate using a mock galaxy catalogue of the
HSC-SSP$\times$unWISE data, in which known identified clusters are
erased. To generate a mock galaxy catalogue, we first remove the
member galaxy candidates within $r_{500}$ of clusters in this work and
in \citet{wh15} from the HSC-SSP$\times$unWISE galaxy catalogue. We
then randomly shuffle the photometric redshifts of the rest galaxies,
but keep the positions of galaxies on the sky unchanged. The mock
catalogue should contain as few real clusters as possible, but have
the same distribution of galaxies on the sky as the original
catalogue. After these steps, we apply the procedures of cluster
identification in Section~\ref{algorithm} with the richness threshold
of $\lambda_{500} \ge 15$. We adopt the detection threshold of ${\rm
  SNR} \ge 4$ for this test. Since no real clusters are present in the
mock galaxy catalogue, any detected ``clusters'' are regarded as false
detections due to the projection effect. The false detection rate is
defined as being the ratio between the number of false detections from
the mock galaxy data and the total number of identified clusters from
the real HSC-SSP$\times$unWISE data.

We obtain ten mock catalogues of galaxies and get an average false
detection rate, which decreases with cluster $\rm SNR$ from about 7\%
for clusters of $4<{\rm SNR}<5$ to about 4\% for clusters of $5<{\rm
  SNR}<6$ and becomes zero for clusters of ${\rm SNR}>8$, as shown in
Fig.~\ref{falserate}. We therefore
set the threshold of ${\rm SNR}\ge5$ for cluster identification in
this paper.

\begin{figure}
  \centering \includegraphics[width = 0.4\textwidth]{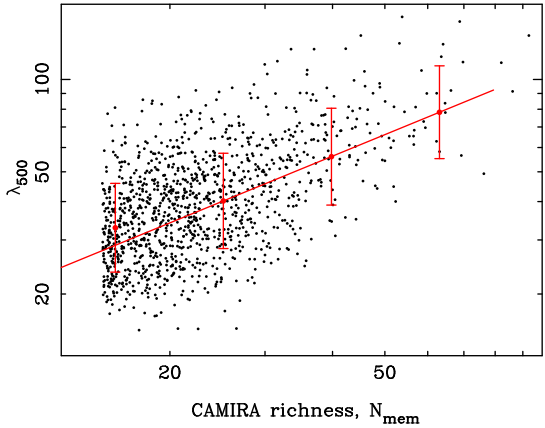}
  \caption{Correlation between the CAMIRA richness \citep{oll+18}
    and the richness of this work. The red dots and error bars are the
    mean and scatter within each bin. The solid is the best fit of
    the correlation.}
\label{compoguri}
\end{figure}

\begin{figure}
  \centering \includegraphics[width = 0.4\textwidth]{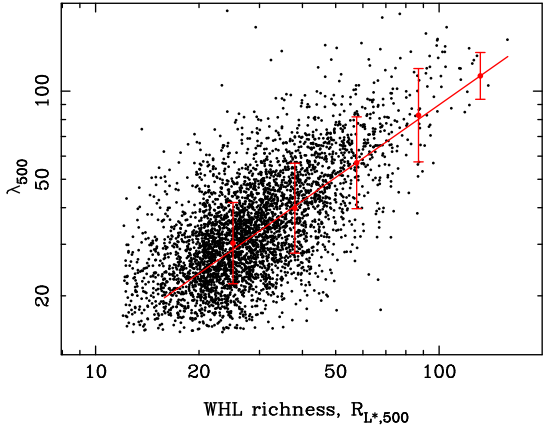}
  \caption{Similar to Fig.~\ref{compoguri} but for the comparison
    with the WHL richness \citep{wh15}.}
\label{compwhl}
\end{figure}

\subsection{Comparison with previous cluster catalogues}
\label{match_cata}

In the HSC-SSP$\times$unWISE region, many galaxy clusters have been
recognized from multiwavelength data, such as optical cluster
catalogues from the Cosmic Evolution Survey (COSMOS) data
\citep{wh11}, the SDSS data \citep{wh15} and the HSC-SSP data
\citep{oll+18}, X-ray cluster catalogue from the XMM-Newton data
\citep{agk+18} and SZ cluster catalogue from the ACT survey
\citep{hhs+18}. We cross-match with these catalogues to investigate
how many known clusters in our cluster catalogue.

\subsubsection{HSC-SSP CAMIRA clusters}

\citet{oll+18} applied the Cluster finding Algorithm based on
Multi-band Identification of Red-sequence gAlaxies
\citep[CAMIRA][]{ogu14} to the HSC-SSP S16A data covering $\sim$232
deg$^2$, and identified 1921 clusters in the redshift range of
$0.1<z<1.1$. The CAMIRA algorithm calculates the likelihood of being
red-sequence galaxies as a function of redshift. The richness is
defined to be the number of red member galaxies with a stellar mass
greater than $10^{10.2}~M_{\odot}$ for cluster candidates, and the
identified clusters have a richness threshold of $N_{\rm mem}\ge 15$
and an equivalent mass of $M_{500}\gtrsim 1.2\times 10^{14}~M_{\odot}$
\citep{cum+20}. The BCG of each cluster was identified as the galaxy
with the highest stellar mass near the richness peak and was taken as
the cluster centre. The uncertainty of cluster redshift is better than
0.01. Simulation shows that the cluster catalogue has a high ($>90\%$)
completeness and purity. The HSC-SSP CAMIRA catalogue was
obtained from the five-band ($grizy$) photometric data, and the
colours are insensitive to redshift for galaxies of $z>1.4$ because
the 4000~\AA~break moves out of the $y$ band. Obviously, we extend the
cluster sample to higher redshifts by combining the HSC-SSP data with
the infrared data of the unWISE (see Fig.~\ref{hist_zc}).

There are 1447 (i.e. 75\%) HSC-SSP CAMIRA clusters within a projected
separation of $1.5\,r_{500}$ and a redshift difference of $0.05(1+z)$
from the clusters in our catalogue. About 70\% of the CAMIRA clusters
with a richness of $N_{\rm mem}\sim 20$ are matched with the clusters
in Table~\ref{tab1}, and about $>90\%$ clusters of $N_{\rm mem}\ge 30$
are matched. The richness of CAMIRA clusters is related to our value
(see Fig.~\ref{compoguri}) by a power law of
\begin{equation}
  \log \lambda_{500}=(0.72\pm0.03)\log N_{\rm mem}+(0.59\pm0.04).
  \label{ogurich}
\end{equation}
In other words, the threshold of CAMIRA richness of $N_{\rm mem}=15$
is equivalent to our richness of $\lambda_{500}\sim27$, suggesting
that the CAMIRA catalogue has a larger mass threshold than our cluster
catalogue. 

\subsubsection{SDSS clusters}

Some of identified clusters from the HSC-SSP$\times$unWISE data have
been detected from the SDSS data \citep[WHL,][]{whl12}. The richness
is defined to be the total luminosity of member galaxies in units of
$L^{\ast}$. \citet{wh15} updated the WHL cluster catalogue with the
spectroscopic redshifts of the SDSS and identified additional 25,419
new clusters at higher redshifts, so that the combined WHL catalogue
contains 158,103 clusters.

In the HSC-SSP$\times$unWISE region, there are 7464 WHL clusters and
4689 (i.e. 63\%) of them are matched with the clusters in
Table~\ref{tab1} within a projected separation of $1.5\,r_{500}$ and a
redshift difference of $0.05(1+z)$, most of which have redshifts
$z<0.65$. About 65\% of these WHL clusters with a richness of $R_{L*,
  500}\sim 30$ are matched and about 85\% of the WHL clusters of
$R_{L*, 500}\sim 50$ are matched. Fig.~\ref{compwhl} shows the
correlation between the richness of this work and the richness in the
WHL catalogue. The best fit to a power law gives
\begin{equation}
\log \lambda_{500}=(0.83\pm0.01)\log R_{L*, 500}+(0.26\pm0.02).
\end{equation}

\begin{figure}
\centering \includegraphics[width = 0.4\textwidth]{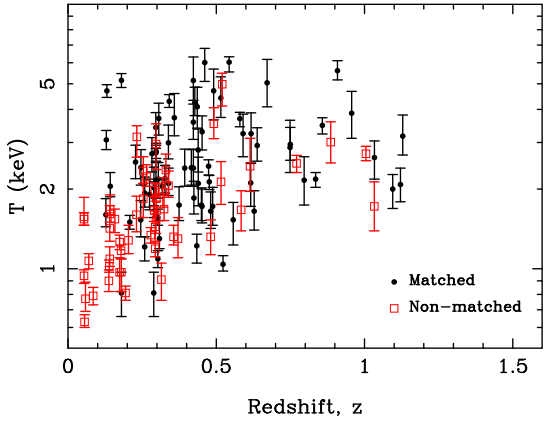}\\
\caption{Matched and non-matched XXL clusters.}
\label{xray}
\end{figure}

\subsubsection{X-ray clusters}

We compare our clusters with those identified from X-ray
observations. A few deep X-ray surveys have been carried out in the
HSC-SSP$\times$unWISE coverage, including XMM-Large-Scale Structure
(LSS) field survey \citep{cal+14}, XMM-COSMOS field survey
\citep{hcb+07} and XMM-XXL field survey \citep{ppa+16}. The XMM-XXL is
among the largest XMM program covering two extragalactic areas of 25
deg$^2$ each with a sensitivity of $\sim 5\times10^{-15}$~erg~s$^{-1}$
in the 0.5--2keV band for point sources. The
XXL cluster catalogue contains 365 clusters at $z<1.2$ \citep{agk+18}.

In the HSC-SSP$\times$unWISE region, we get 122 XXL clusters with
temperature data, of which 75 clusters are matched by our
cluster catalogue. Fig.~\ref{xray} shows the XXL clusters in the
temperature--redshift plane. About 80\% of XXL clusters of
$T=2$--$4.5$ keV and about 90\% of clusters of $T>4.5$ keV are found
in the Table~\ref{tab1}.

\begin{figure}
\centering \includegraphics[width = 0.4\textwidth]{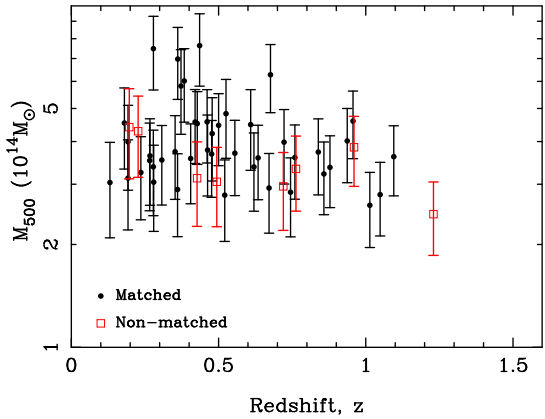}
\caption{Matched and non-matched ACT clusters.}
\label{act}
\end{figure}

\subsubsection{SZ clusters}

The SZ effect has the advantage to detect high-redshift clusters
because of its redshift-independent brightness. The ACT survey has
been carried out in the HSC-SSP$\times$unWISE coverage to identify
high-redshift SZ clusters \citep{hhm+13}. The latest ACT cluster
catalogue includes 182 SZ clusters in the redshift range of
$0.1<z<1.4$ from the 987.5 deg$^2$ celestial equator field
\citep{hhs+18}. There are 52 ACT clusters in the HSC-SSP$\times$unWISE
region, of which 44 (85\%) clusters are matched by our cluster
catalogue (Fig.~\ref{act}), and eight non-matched clusters have a lower
SNR or a lower richness than the threshold for cluster identification
by our algorithm.

\subsubsection{Clusters in the COSMOS field}
\label{cosmos}

The COSMOS is designed to study evolution of galaxies, active galactic
nuclei and large scale structure over the redshift range of $0.5<z<6$
using deep imaging and spectroscopic observations \citep{sab+07},
covering an area of 2
deg$^2$ at the wavelengths of near-ultraviolet \citep[NUV; e.g.][]{zsr+07}, optical
\citep[e.g.][]{saa+07,tsm+07}, infrared \citep[e.g.][]{ssa+07} and
X-ray \citep[e.g.][]{hcb+07,cmc+16}. Clusters/groups have been
identified using spectroscopic data
\citep[e.g.][]{kli+12}, the 30-band NUV--infrared photometric data
\citep[e.g.][]{zrw+07,bmh+11,wh11} and X-ray imaging data
\citep[e.g.][]{fgh+07}.
In our catalogue, there are 72 clusters in the COSMOS field, of which
68 clusters have been previously detected
\citep{fgh+07,zrw+07,gbm09,wh11,whl12,scc+12,afh+12,bmm+14}, and four
clusters are newly identified at $z>0.8$. Such a high matched
rate suggests that our clusters is fairly reliable, verifying the low
false detection rate based on the Monte-Carlo simulation in
Section~\ref{false}.

\section{Properties of member galaxies}

Using identified clusters in a wide redshift range from the
HSC-SSP$\times$unWISE data, we study the evolution of BCG stellar mass
and star formation.

\subsection{Stellar mass evolution of BCGs}

\begin{figure}
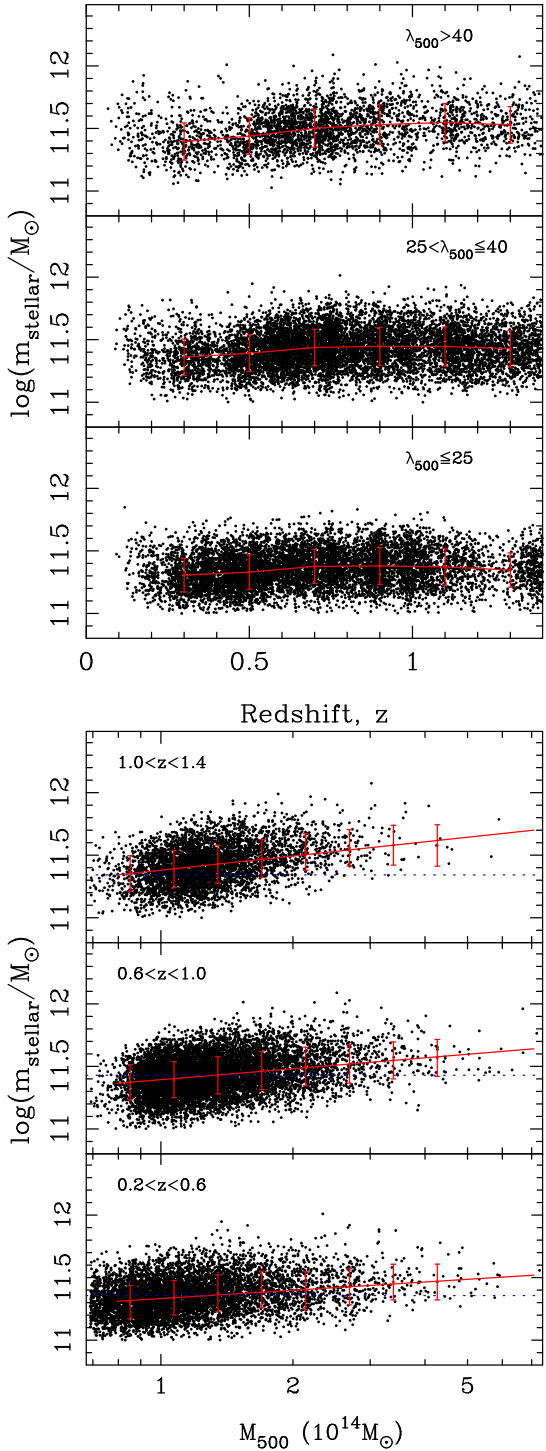

\centering
\includegraphics[width=0.4\textwidth]{f17a.eps} %\hspace{3mm}
\includegraphics[width=0.4\textwidth]{f17b.eps}
\caption{Upper: BCG stellar mass as a function of redshift within three
  richness bins.  Lower: correlation between BCG stellar mass and
  cluster mass within three redshift bins. The red solid lines are the
  best fit to the data.}
\label{bcg2}
\end{figure}

According to the hierarchical model of galaxy formation, the
population of BCGs was formed at early universe and grew through the
merging of satellite galaxies \citep{db07}. However, no significant
evolution of the BCG stellar mass is shown in optical selected
clusters since $z\sim1$ \citep{wad+08}, probably due to the selection
effect because BCGs co-evolve with their host clusters. \citet{lsm+12}
found that the BCG stellar mass increases by a factor of 1.8$\pm$0.3
from $z=0.9$ to $z=0.2$, but the growth is much slower than the
prediction of the semi-analytic model in \citet{db07}. \citet{blm+16}
argued that the growth of BCG stellar mass is consistent with the
model.
In addition, the BCG stellar mass is related to cluster
mass. \citet{zmm+16} found the scaling relation of $m_{\rm stellar}\propto
M_{200}^{0.24\pm0.08}(1+z)^{-0.19\pm0.34}$ using 106 X-ray selected
cluster of $z<1.2$. \citet{eff+19} found from 416 BCGs that the
dependence of stellar masses on halo mass does not significantly
change within $0.1<z<0.65$. The slope is 0.41$\pm$0.04 within $0.1\le z\le
0.3$ and 0.31$\pm$0.02 within $0.3<z\le 0.65$. Using 42 groups and
clusters at $0.05<z<1.75$, \citet{dgz+20} found the scaling relation
of $m_{\rm stellar}\propto M_{500}^{0.48\pm0.06}$ for the BCG plus
intracluster light. The normalization does not change from $z=0.1$ to
$z=0.4$, but is 2.08$\pm$0.21 times higher than that at $z=1.55$.

Using the large number of clusters in a wide redshift range, we can
investigate the evolution of BCG stellar mass with redshift as well as
the dependence on cluster mass. First, we divide our cluster sample
into three richness bins and study the redshift evolution of BCG
stellar mass. Within each richness bin, the BCG stellar mass has a
scatter of 0.14 dex and does not evolve with redshift (upper panel of
Fig.~\ref{bcg2}). Then, we divide our cluster sample into three
redshift bins and study the dependence of BCG stellar mass on cluster
mass. We find that the BCG stellar mass is positively related to
cluster mass (lower panels of Fig.~\ref{bcg2}) by the relations
\begin{eqnarray}
\log m_{\rm stellar}=(0.23\pm0.02)\log M_{500}+(11.33\pm0.01) \nonumber \\
{\rm for}~ 0.2<z<0.6; \nonumber \\
\log m_{\rm stellar}=(0.29\pm0.01)\log M_{500}+(11.39\pm0.01) \nonumber \\
{\rm for}~ 0.6<z<1.0; \nonumber \\
\log m_{\rm stellar}=(0.38\pm0.03)\log M_{500}+(11.38\pm0.02) \nonumber \\
     {\rm for}~ 1.0<z<1.4,
\label{bcgcluster}
\end{eqnarray}
which implies no significant evolution with redshift but dependence on
cluster environments. The slope we obtain is consistent with that of
\citet{zmm+16}, but slightly lower than those in \citet{eff+19} and
\citet{dgz+20} and also that of 0.4--0.5 from the theoretical model
\citep{wad+08}.

Because of the BCG--cluster co-evolution, the results shown in
Fig.~\ref{bcg2} do not mean that the BCG stellar mass is constant over
cosmic time because cluster mass may evolve with redshift. The upper
panel of Fig.~\ref{bcg2} is obtained from the given range of cluster
richness. According to the formula of halo merger rate \citep{fmb10},
the clusters are about three times more massive at $z=1$ than those at
$z=0$. Using the Eq.~(\ref{bcgcluster}) (adopting an average slope
of 0.3), we find that the BCG stellar mass increases by a factor of
1.5 from the redshift of $z=1$ to $z=0$, consistent with the suggestions
by \citet{lsm+12} and \citet{zmm+16}, but lower than a factor of 3 by
the semi-analytic model \citep{db07}.

\begin{figure}
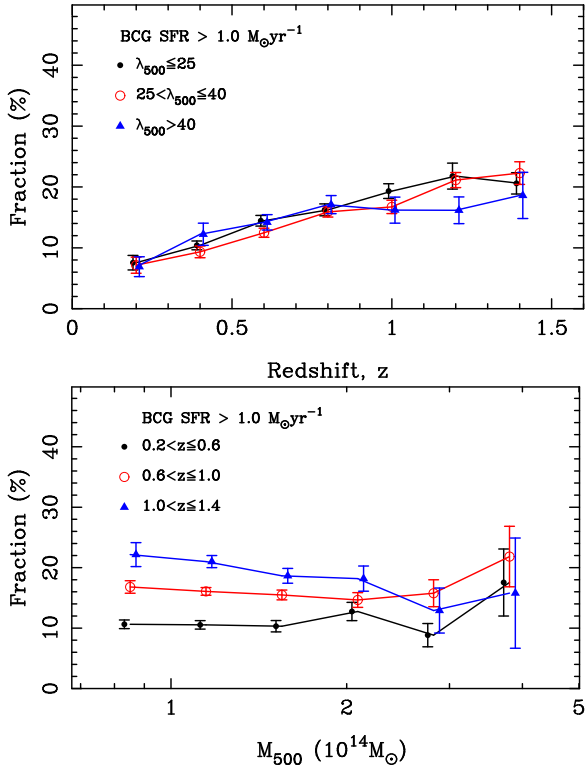

\centering
\includegraphics[width=0.43\textwidth]{f18a.eps}
\includegraphics[width=0.43\textwidth]{f18b.eps}
\caption{Fraction of star-forming BCGs as a function of redshift
  (upper) and cluster mass (lower).}
\label{bcgsfr}
\end{figure}

\subsection{Star formation in BCGs}

Optical and infrared data show that the stellar population in most
BCGs is in agreement with a passive evolution. However, BCGs in some
cool-core clusters have signatures of ongoing star formation
\citep{cae+99,lmm12}, though the fraction of star-forming BCGs is
often low.

The star formation rates of the BCGs can be obtained by fitting the
7-band photometry to the Fitting and Assessment of Synthetic Templates
code \citep[FAST;][]{kvl+09}. We adopt the stellar population
synthesis models from \citet{bc03} and the initial mass function of
\citet{cha03}, and also an exponentially declining star formation
history. We calculate the fraction of BCGs with a star formation rate
greater than 1.0 $M_{\odot}$~yr$^{-1}$ for the clusters of different
redshift bins and richness bins. As shown in the upper panel of
Fig.~\ref{bcgsfr}, the fraction increases with redshift for all
three richness bins, from about 8\% at $z\sim 0.2$ to 20\% at $z\sim
1.4$. In the lower panel of Fig.~\ref{bcgsfr}, we find that the
fraction of star-forming BCGs does not significantly vary with
cluster mass.
Previously, \citet{coc19} found that a fraction of 9\% of BCGs are
star forming at $z\sim0.2$ based on the SDSS and WISE
data. \citet{wmn+15} presented a 24 $\mu$m study of 535 BCGs of
$0.2<z<1.8$ from the Spitzer Adaptation of the Red-Sequence Cluster
Survey. They found that $\sim$20\% of BCGs at $z>1$ have a high
infrared luminosity of $L_{IR}>10^{12}~L_{\odot}$ powered by star
formation. Our results in Fig.~\ref{bcgsfr} are consistent with these
conclusions.

\section{Summary}

We obtain a catalogue of photometric redshifts for 14.68 million
HSC-SSP$\times$unWISE galaxies. The photometric redshifts of galaxies
are estimated in the colour space based on the nearest-neighbour
algorithm. The redshift uncertainty is about 0.024 at redshifts
$z<0.7$ and increases with redshift to about 0.11 at
$z\sim2$. Comparing to the public HSC-SSP redshifts, we improve the
accuracy of photometric redshifts for galaxies within $1.4<z<2$. The
stellar masses of galaxies are derived from the infrared magnitude
and calibrated using the COSMOS data. The HSC-SSP$\times$unWISE
catalogue is $>90\%$ complete at $z<0.5$ for galaxies with stellar
masses $>10^{10.5}~M_{\odot}$ and becomes $\sim70\%$ complete at
$z\sim1$ .

From this large photometric redshift catalogue, we then identify
21,661 clusters in the redshift range of $0.1<z\lesssim2$ with a
detection signal-to-noise ratio $\ge5$ and an equivalent mass
$M_{500}\ge0.7\times10^{14}~M_{\odot}$. The uncertainty of cluster
redshift is about $0.022$. Simulations show that the completeness is
more than 90\% for massive clusters of
$M_{500}>2\times10^{14}~M_{\odot}$ at $z<1$ and the false detection is
less than 5\%.
We study the stellar mass and star formation rate of the
BCGs. We find no significant redshift evolution for the scaling
relation between BCG stellar mass and cluster mass. The fraction of
star-forming BCGs increases with redshift, but does not depend on
cluster mass. 

\section*{Acknowledgements}

We thank the referee for valuable comments that helped to improve the
paper. We are grateful to Dr. F. S. Liu for discussions.
Dr. L. G. Hou and X. M. Meng help to figure out software. The authors
are partially supported by the National Natural Science Foundation of
China (Grant No. 11988101 and U1731127), the Key Research Program of
the Chinese Academy of Sciences (Grant No. QYZDJ-SSW-SLH021) and the
strategic Priority Research Program of Chinese Academy of Sciences
(Grant No. XDB23010200), and the Open Project Program of the Key
Laboratory of FAST, NAOC, Chinese Academy of Sciences.
The HSC collaboration includes the astronomical
communities of Japan and Taiwan, and Princeton University. The HSC
instrumentation and software were developed by the National
Astronomical Observatory of Japan (NAOJ), the Kavli Institute for the
Physics and Mathematics of the Universe (Kavli IPMU), the University
of Tokyo, the High Energy Accelerator Research Organization (KEK), the
Academia Sinica Institute for Astronomy and Astrophysics in Taiwan
(ASIAA), and Princeton University. Funding was contributed by the
FIRST program from Japanese Cabinet Office, the Ministry of Education,
Culture, Sports, Science and Technology (MEXT), the Japan Society for
the Promotion of Science (JSPS), Japan Science and Technology Agency
(JST), the Toray Science Foundation, NAOJ, Kavli IPMU, KEK, ASIAA, and
Princeton University.

This paper has used software developed for the Large Synoptic
Survey Telescope. We thank the LSST Project for making their code
available as free software at http://dm.lsst.org

This paper is based [in part] on data collected at the Subaru
Telescope and retrieved from the HSC data archive system, which is
operated by Subaru Telescope and Astronomy Data Center at NAOJ. Data
analysis was in part carried out with the cooperation of Center for
Computational Astrophysics, NAOJ.

This publication has used data products from the Wide-field
Infrared Survey Explorer, which is a joint project of the University
of California, Los Angeles, and the Jet Propulsion
Laboratory/California Institute of Technology, funded by the National
Aeronautics and Space Administration.

\section*{Data availability}

The data of 14.68 million HSC-SSP$\times$unWISE galaxies (7-band
magnitudes, photometric redshifts and stellar masses), 21,661
identified clusters and their member galaxy candidates are publicly
available at http://zmtt.bao.ac.cn/galaxy\_clusters/.

\bibliographystyle{mnras}
%\bibliography{journals,wise}
\bibliography{wise}

\begin{thebibliography}{}
\makeatletter
\relax
\def\mn@urlcharsother{\let\do\@makeother \do\$\do\&\do\#\do\^\do\_\do\%\do\~}
\def\mn@doi{\begingroup\mn@urlcharsother \@ifnextchar [ {\mn@doi@}
  {\mn@doi@[]}}
\def\mn@doi@[#1]#2{\def\@tempa{#1}\ifx\@tempa\@empty \href
  {http://dx.doi.org/#2} {doi:#2}\else \href {http://dx.doi.org/#2} {#1}\fi
  \endgroup}
\def\mn@eprint#1#2{\mn@eprint@#1:#2::\@nil}
\def\mn@eprint@arXiv#1{\href {http://arxiv.org/abs/#1} {{\tt arXiv:#1}}}
\def\mn@eprint@dblp#1{\href {http://dblp.uni-trier.de/rec/bibtex/#1.xml}
  {dblp:#1}}
\def\mn@eprint@#1:#2:#3:#4\@nil{\def\@tempa {#1}\def\@tempb {#2}\def\@tempc
  {#3}\ifx \@tempc \@empty \let \@tempc \@tempb \let \@tempb \@tempa \fi \ifx
  \@tempb \@empty \def\@tempb {arXiv}\fi \@ifundefined
  {mn@eprint@\@tempb}{\@tempb:\@tempc}{\expandafter \expandafter \csname
  mn@eprint@\@tempb\endcsname \expandafter{\@tempc}}}

\bibitem[\protect\citeauthoryear{{Abell}}{{Abell}}{1958}]{abe58}
{Abell} G.~O.,  1958, \apjs, \href
  {http://adsabs.harvard.edu/abs/1958ApJS....3..211A} {3, 211}

\bibitem[\protect\citeauthoryear{{Abell}, {Corwin}  \& {Olowin}}{{Abell}
  et~al.}{1989}]{aco89}
{Abell} G.~O.,  {Corwin} Jr. H.~G.,   {Olowin} R.~P.,  1989, \mn@doi [\apjs]
  {10.1086/191333}, \href {http://adsabs.harvard.edu/abs/1989ApJS...70....1A}
  {70, 1}

\bibitem[\protect\citeauthoryear{{Abolfathi} et~al.,}{{Abolfathi}
  et~al.}{2018}]{dr14+18}
{Abolfathi} B.,  et~al., 2018, \mn@doi [\apjs] {10.3847/1538-4365/aa9e8a},
  \href {https://ui.adsabs.harvard.edu/abs/2018ApJS..235...42A} {235, 42}

\bibitem[\protect\citeauthoryear{{Adami} et~al.,}{{Adami}
  et~al.}{2018}]{agk+18}
{Adami} C.,  et~al., 2018, \mn@doi [\aap] {10.1051/0004-6361/201731606}, \href
  {https://ui.adsabs.harvard.edu/abs/2018A%26A...620A...5A} {620, A5}

\bibitem[\protect\citeauthoryear{{Aihara} et~al.,}{{Aihara}
  et~al.}{2018}]{hscssp18}
{Aihara} H.,  et~al., 2018, \mn@doi [\pasj] {10.1093/pasj/psx066}, \href
  {https://ui.adsabs.harvard.edu/abs/2018PASJ...70S...4A} {70, S4}

\bibitem[\protect\citeauthoryear{{Aihara} et~al.,}{{Aihara}
  et~al.}{2019}]{hscdr2}
{Aihara} H.,  et~al., 2019, \mn@doi [\pasj] {10.1093/pasj/psz103}, \href
  {https://ui.adsabs.harvard.edu/abs/2019PASJ..tmp..106A} {p.~106}

\bibitem[\protect\citeauthoryear{{Allen}, {Rapetti}, {Schmidt}, {Ebeling},
  {Morris}  \& {Fabian}}{{Allen} et~al.}{2008}]{ars+08}
{Allen} S.~W.,  {Rapetti} D.~A.,  {Schmidt} R.~W.,  {Ebeling} H.,  {Morris}
  R.~G.,   {Fabian} A.~C.,  2008, \mn@doi [\mnras]
  {10.1111/j.1365-2966.2007.12610.x}, \href
  {http://adsabs.harvard.edu/abs/2008MNRAS.383..879A} {383, 879}

\bibitem[\protect\citeauthoryear{{Allen}, {Evrard}  \& {Mantz}}{{Allen}
  et~al.}{2011}]{aem11}
{Allen} S.~W.,  {Evrard} A.~E.,   {Mantz} A.~B.,  2011, \mn@doi [\araa]
  {10.1146/annurev-astro-081710-102514}, \href
  {http://adsabs.harvard.edu/abs/2011ARA%26A..49..409A} {49, 409}

\bibitem[\protect\citeauthoryear{{Allevato} et~al.,}{{Allevato}
  et~al.}{2012}]{afh+12}
{Allevato} V.,  et~al., 2012, \mn@doi [\apj] {10.1088/0004-637X/758/1/47},
  \href {https://ui.adsabs.harvard.edu/abs/2012ApJ...758...47A} {758, 47}

\bibitem[\protect\citeauthoryear{{Balogh} et~al.,}{{Balogh}
  et~al.}{2014}]{bmm+14}
{Balogh} M.~L.,  et~al., 2014, \mn@doi [\mnras] {10.1093/mnras/stu1332}, \href
  {https://ui.adsabs.harvard.edu/abs/2014MNRAS.443.2679B} {443, 2679}

\bibitem[\protect\citeauthoryear{{Banerjee}, {Szabo}, {Pierpaoli}, {Franco},
  {Ortiz}, {Oramas}  \& {Tornello}}{{Banerjee} et~al.}{2018}]{bsp+18}
{Banerjee} P.,  {Szabo} T.,  {Pierpaoli} E.,  {Franco} G.,  {Ortiz} M.,
  {Oramas} A.,   {Tornello} B.,  2018, \mn@doi [\na]
  {10.1016/j.newast.2017.07.008}, \href
  {http://adsabs.harvard.edu/abs/2018NewA...58...61B} {58, 61}

\bibitem[\protect\citeauthoryear{{Bartelmann}}{{Bartelmann}}{1996}]{bar96}
{Bartelmann} M.,  1996, \aap, \href
  {http://adsabs.harvard.edu/abs/1996A%26A...313..697B} {313, 697}

\bibitem[\protect\citeauthoryear{{Beck}, {Dobos}, {Budav{\'a}ri}, {Szalay}  \&
  {Csabai}}{{Beck} et~al.}{2016}]{brd+16}
{Beck} R.,  {Dobos} L.,  {Budav{\'a}ri} T.,  {Szalay} A.~S.,   {Csabai} I.,
  2016, \mn@doi [\mnras] {10.1093/mnras/stw1009}, \href
  {https://ui.adsabs.harvard.edu/abs/2016MNRAS.460.1371B} {460, 1371}

\bibitem[\protect\citeauthoryear{{Bell} \& {de Jong}}{{Bell} \& {de
  Jong}}{2001}]{bd01}
{Bell} E.~F.,  {de Jong} R.~S.,  2001, \mn@doi [\apj] {10.1086/319728}, \href
  {https://ui.adsabs.harvard.edu/abs/2001ApJ...550..212B} {550, 212}

\bibitem[\protect\citeauthoryear{{Bellagamba}, {Maturi}, {Hamana},
  {Meneghetti}, {Miyazaki}  \& {Moscardini}}{{Bellagamba}
  et~al.}{2011}]{bmh+11}
{Bellagamba} F.,  {Maturi} M.,  {Hamana} T.,  {Meneghetti} M.,  {Miyazaki} S.,
   {Moscardini} L.,  2011, \mn@doi [\mnras] {10.1111/j.1365-2966.2011.18202.x},
  \href {https://ui.adsabs.harvard.edu/abs/2011MNRAS.413.1145B} {413, 1145}

\bibitem[\protect\citeauthoryear{{Bellstedt} et~al.,}{{Bellstedt}
  et~al.}{2016}]{blm+16}
{Bellstedt} S.,  et~al., 2016, \mn@doi [\mnras] {10.1093/mnras/stw1184}, \href
  {http://adsabs.harvard.edu/abs/2016MNRAS.460.2862B} {460, 2862}

\bibitem[\protect\citeauthoryear{{Bernardi}, {Shankar}, {Hyde}, {Mei},
  {Marulli}  \& {Sheth}}{{Bernardi} et~al.}{2010}]{bsh+10}
{Bernardi} M.,  {Shankar} F.,  {Hyde} J.~B.,  {Mei} S.,  {Marulli} F.,
  {Sheth} R.~K.,  2010, \mn@doi [\mnras] {10.1111/j.1365-2966.2010.16425.x},
  \href {https://ui.adsabs.harvard.edu/abs/2010MNRAS.404.2087B} {404, 2087}

\bibitem[\protect\citeauthoryear{{Bilicki} et~al.,}{{Bilicki}
  et~al.}{2016}]{bpj+16}
{Bilicki} M.,  et~al., 2016, \mn@doi [\apjs] {10.3847/0067-0049/225/1/5}, \href
  {http://adsabs.harvard.edu/abs/2016ApJS..225....5B} {225, 5}

\bibitem[\protect\citeauthoryear{{Binggeli}, {Tammann}  \&
  {Sandage}}{{Binggeli} et~al.}{1987}]{bts87}
{Binggeli} B.,  {Tammann} G.~A.,   {Sandage} A.,  1987, \mn@doi [\aj]
  {10.1086/114467}, \href {http://adsabs.harvard.edu/abs/1987AJ.....94..251B}
  {94, 251}

\bibitem[\protect\citeauthoryear{{B{\"o}hringer}, {Chon}, {Collins}, {Guzzo},
  {Nowak}  \& {Bobrovskyi}}{{B{\"o}hringer} et~al.}{2013}]{bcc+13}
{B{\"o}hringer} H.,  {Chon} G.,  {Collins} C.~A.,  {Guzzo} L.,  {Nowak} N.,
  {Bobrovskyi} S.,  2013, \mn@doi [\aap] {10.1051/0004-6361/201220722}, \href
  {http://adsabs.harvard.edu/abs/2013A%26A...555A..30B} {555, A30}

\bibitem[\protect\citeauthoryear{{B{\"o}hringer}, {Chon}, {Retzlaff},
  {Tr{\"u}mper}, {Meisenheimer}  \& {Schartel}}{{B{\"o}hringer}
  et~al.}{2017}]{bcr+17}
{B{\"o}hringer} H.,  {Chon} G.,  {Retzlaff} J.,  {Tr{\"u}mper} J.,
  {Meisenheimer} K.,   {Schartel} N.,  2017, \mn@doi [\aj]
  {10.3847/1538-3881/aa67ed}, \href
  {http://adsabs.harvard.edu/abs/2017AJ....153..220B} {153, 220}

\bibitem[\protect\citeauthoryear{{Bolzonella}, {Miralles}  \&
  {Pell{\'o}}}{{Bolzonella} et~al.}{2000}]{bmp00}
{Bolzonella} M.,  {Miralles} J.~M.,   {Pell{\'o}} R.,  2000, \aap, \href
  {https://ui.adsabs.harvard.edu/abs/2000A&A...363..476B} {363, 476}

\bibitem[\protect\citeauthoryear{{Bradshaw} et~al.,}{{Bradshaw}
  et~al.}{2013}]{bah+13}
{Bradshaw} E.~J.,  et~al., 2013, \mn@doi [\mnras] {10.1093/mnras/stt715}, \href
  {https://ui.adsabs.harvard.edu/abs/2013MNRAS.433..194B} {433, 194}

\bibitem[\protect\citeauthoryear{{Brammer}, {van Dokkum}  \& {Coppi}}{{Brammer}
  et~al.}{2008}]{bvc08}
{Brammer} G.~B.,  {van Dokkum} P.~G.,   {Coppi} P.,  2008, \mn@doi [\apj]
  {10.1086/591786}, \href
  {https://ui.adsabs.harvard.edu/abs/2008ApJ...686.1503B} {686, 1503}

\bibitem[\protect\citeauthoryear{{Brodwin} et~al.,}{{Brodwin}
  et~al.}{2012}]{bgs+12}
{Brodwin} M.,  et~al., 2012, \mn@doi [\apj] {10.1088/0004-637X/753/2/162},
  \href {https://ui.adsabs.harvard.edu/abs/2012ApJ...753..162B} {753, 162}

\bibitem[\protect\citeauthoryear{{Bruzual} \& {Charlot}}{{Bruzual} \&
  {Charlot}}{2003}]{bc03}
{Bruzual} G.,  {Charlot} S.,  2003, \mn@doi [\mnras]
  {10.1046/j.1365-8711.2003.06897.x}, \href
  {http://adsabs.harvard.edu/abs/2003MNRAS.344.1000B} {344, 1000}

\bibitem[\protect\citeauthoryear{{Butcher} \& {Oemler}}{{Butcher} \&
  {Oemler}}{1978}]{bo78}
{Butcher} H.,  {Oemler} Jr. A.,  1978, \mn@doi [\apj] {10.1086/155751}, \href
  {http://ads.ari.uni-heidelberg.de/abs/1978ApJ...219...18B} {219, 18}

\bibitem[\protect\citeauthoryear{{Butcher} \& {Oemler}}{{Butcher} \&
  {Oemler}}{1984}]{bo84}
{Butcher} H.,  {Oemler} Jr. A.,  1984, \mn@doi [\apj] {10.1086/162519}, \href
  {http://ads.ari.uni-heidelberg.de/abs/1984ApJ...285..426B} {285, 426}

\bibitem[\protect\citeauthoryear{{Carrasco Kind} \& {Brunner}}{{Carrasco Kind}
  \& {Brunner}}{2014}]{cb14}
{Carrasco Kind} M.,  {Brunner} R.~J.,  2014, \mn@doi [\mnras]
  {10.1093/mnras/stt2456}, \href
  {https://ui.adsabs.harvard.edu/abs/2014MNRAS.438.3409C} {438, 3409}

\bibitem[\protect\citeauthoryear{{Cerulo}, {Orellana}  \& {Covone}}{{Cerulo}
  et~al.}{2019}]{coc19}
{Cerulo} P.,  {Orellana} G.~A.,   {Covone} G.,  2019, \mn@doi [\mnras]
  {10.1093/mnras/stz1495}, \href
  {https://ui.adsabs.harvard.edu/abs/2019MNRAS.487.3759C} {487, 3759}

\bibitem[\protect\citeauthoryear{{Chabrier}}{{Chabrier}}{2003}]{cha03}
{Chabrier} G.,  2003, \mn@doi [\pasp] {10.1086/376392}, \href
  {http://adsabs.harvard.edu/abs/2003PASP..115..763C} {115, 763}

\bibitem[\protect\citeauthoryear{{Chiu}, {Umetsu}, {Murata}, {Medezinski}  \&
  {Oguri}}{{Chiu} et~al.}{2020}]{cum+20}
{Chiu} I.~N.,  {Umetsu} K.,  {Murata} R.,  {Medezinski} E.,   {Oguri} M.,
  2020, \mn@doi [\mnras] {10.1093/mnras/staa1158}, \href
  {https://ui.adsabs.harvard.edu/abs/2020MNRAS.495..428C} {495, 428}

\bibitem[\protect\citeauthoryear{{Civano} et~al.,}{{Civano}
  et~al.}{2016}]{cmc+16}
{Civano} F.,  et~al., 2016, \mn@doi [\apj] {10.3847/0004-637X/819/1/62}, \href
  {https://ui.adsabs.harvard.edu/abs/2016ApJ...819...62C} {819, 62}

\bibitem[\protect\citeauthoryear{{Clerc} et~al.,}{{Clerc}
  et~al.}{2014}]{cal+14}
{Clerc} N.,  et~al., 2014, \mn@doi [\mnras] {10.1093/mnras/stu1625}, \href
  {http://adsabs.harvard.edu/abs/2014MNRAS.444.2723C} {444, 2723}

\bibitem[\protect\citeauthoryear{{Cluver} et~al.,}{{Cluver}
  et~al.}{2014}]{cjh+14}
{Cluver} M.~E.,  et~al., 2014, \mn@doi [\apj] {10.1088/0004-637X/782/2/90},
  \href {https://ui.adsabs.harvard.edu/abs/2014ApJ...782...90C} {782, 90}

\bibitem[\protect\citeauthoryear{{Colberg}, {White}, {Jenkins}  \&
  {Pearce}}{{Colberg} et~al.}{1999}]{cwj+99}
{Colberg} J.~M.,  {White} S.~D.~M.,  {Jenkins} A.,   {Pearce} F.~R.,  1999,
  \mn@doi [\mnras] {10.1046/j.1365-8711.1999.02400.x}, \href
  {http://adsabs.harvard.edu/abs/1999MNRAS.308..593C} {308, 593}

\bibitem[\protect\citeauthoryear{{Collister} \& {Lahav}}{{Collister} \&
  {Lahav}}{2004}]{cl04}
{Collister} A.~A.,  {Lahav} O.,  2004, \mn@doi [\pasp] {10.1086/383254}, \href
  {https://ui.adsabs.harvard.edu/abs/2004PASP..116..345C} {116, 345}

\bibitem[\protect\citeauthoryear{{Cool} et~al.,}{{Cool} et~al.}{2013}]{cmb+13}
{Cool} R.~J.,  et~al., 2013, \mn@doi [\apj] {10.1088/0004-637X/767/2/118},
  \href {https://ui.adsabs.harvard.edu/abs/2013ApJ...767..118C} {767, 118}

\bibitem[\protect\citeauthoryear{{Cooper} et~al.,}{{Cooper}
  et~al.}{2011}]{cac+11}
{Cooper} M.~C.,  et~al., 2011, \mn@doi [\apjs] {10.1088/0067-0049/193/1/14},
  \href {https://ui.adsabs.harvard.edu/abs/2011ApJS..193...14C} {193, 14}

\bibitem[\protect\citeauthoryear{{Crawford}, {Allen}, {Ebeling}, {Edge}  \&
  {Fabian}}{{Crawford} et~al.}{1999}]{cae+99}
{Crawford} C.~S.,  {Allen} S.~W.,  {Ebeling} H.,  {Edge} A.~C.,   {Fabian}
  A.~C.,  1999, \mn@doi [\mnras] {10.1046/j.1365-8711.1999.02583.x}, \href
  {http://adsabs.harvard.edu/abs/1999MNRAS.306..857C} {306, 857}

\bibitem[\protect\citeauthoryear{{Cunha}, {Lima}, {Oyaizu}, {Frieman}  \&
  {Lin}}{{Cunha} et~al.}{2009}]{clo+09}
{Cunha} C.~E.,  {Lima} M.,  {Oyaizu} H.,  {Frieman} J.,   {Lin} H.,  2009,
  \mn@doi [\mnras] {10.1111/j.1365-2966.2009.14908.x}, \href
  {https://ui.adsabs.harvard.edu/abs/2009MNRAS.396.2379C} {396, 2379}

\bibitem[\protect\citeauthoryear{{Cutri} \& {et al.}}{{Cutri} \& {et
  al.}}{2013}]{allwise13}
{Cutri} R.~M.,  {et al.} 2013, VizieR Online Data Catalog, \href
  {https://ui.adsabs.harvard.edu/abs/2013yCat.2328....0C} {2328}

\bibitem[\protect\citeauthoryear{{De Lucia} \& {Blaizot}}{{De Lucia} \&
  {Blaizot}}{2007}]{db07}
{De Lucia} G.,  {Blaizot} J.,  2007, \mn@doi [\mnras]
  {10.1111/j.1365-2966.2006.11287.x}, \href
  {http://adsabs.harvard.edu/abs/2007MNRAS.375....2D} {375, 2}

\bibitem[\protect\citeauthoryear{{DeMaio} et~al.,}{{DeMaio}
  et~al.}{2020}]{dgz+20}
{DeMaio} T.,  et~al., 2020, \mn@doi [\mnras] {10.1093/mnras/stz3236}, \href
  {https://ui.adsabs.harvard.edu/abs/2020MNRAS.491.3751D} {491, 3751}

\bibitem[\protect\citeauthoryear{{Drinkwater} et~al.,}{{Drinkwater}
  et~al.}{2010}]{djb+10}
{Drinkwater} M.~J.,  et~al., 2010, \mn@doi [\mnras]
  {10.1111/j.1365-2966.2009.15754.x}, \href
  {https://ui.adsabs.harvard.edu/abs/2010MNRAS.401.1429D} {401, 1429}

\bibitem[\protect\citeauthoryear{{Drory}, {Bender}  \& {Hopp}}{{Drory}
  et~al.}{2004}]{dbh04}
{Drory} N.,  {Bender} R.,   {Hopp} U.,  2004, \mn@doi [\apjl] {10.1086/426502},
  \href {https://ui.adsabs.harvard.edu/abs/2004ApJ...616L.103D} {616, L103}

\bibitem[\protect\citeauthoryear{{Eisenhardt} et~al.,}{{Eisenhardt}
  et~al.}{2008}]{ebg+08}
{Eisenhardt} P.~R.~M.,  et~al., 2008, \mn@doi [\apj] {10.1086/590105}, \href
  {http://adsabs.harvard.edu/abs/2008ApJ...684..905E} {684, 905}

\bibitem[\protect\citeauthoryear{{Erfanianfar} et~al.,}{{Erfanianfar}
  et~al.}{2013}]{eft+13}
{Erfanianfar} G.,  et~al., 2013, \mn@doi [\apj] {10.1088/0004-637X/765/2/117},
  \href {https://ui.adsabs.harvard.edu/abs/2013ApJ...765..117E} {765, 117}

\bibitem[\protect\citeauthoryear{{Erfanianfar} et~al.,}{{Erfanianfar}
  et~al.}{2019}]{eff+19}
{Erfanianfar} G.,  et~al., 2019, \mn@doi [\aap] {10.1051/0004-6361/201935375},
  \href {https://ui.adsabs.harvard.edu/abs/2019A&A...631A.175E} {631, A175}

\bibitem[\protect\citeauthoryear{{Fakhouri}, {Ma}  \&
  {Boylan-Kolchin}}{{Fakhouri} et~al.}{2010}]{fmb10}
{Fakhouri} O.,  {Ma} C.-P.,   {Boylan-Kolchin} M.,  2010, \mn@doi [\mnras]
  {10.1111/j.1365-2966.2010.16859.x}, \href
  {https://ui.adsabs.harvard.edu/abs/2010MNRAS.406.2267F} {406, 2267}

\bibitem[\protect\citeauthoryear{{Finoguenov} et~al.,}{{Finoguenov}
  et~al.}{2007}]{fgh+07}
{Finoguenov} A.,  et~al., 2007, \mn@doi [\apjs] {10.1086/516577}, \href
  {https://ui.adsabs.harvard.edu/abs/2007ApJS..172..182F} {172, 182}

\bibitem[\protect\citeauthoryear{{Garilli} et~al.,}{{Garilli}
  et~al.}{2014}]{ggs+14}
{Garilli} B.,  et~al., 2014, \mn@doi [\aap] {10.1051/0004-6361/201322790},
  \href {https://ui.adsabs.harvard.edu/abs/2014A%26A...562A..23G} {562, A23}

\bibitem[\protect\citeauthoryear{{Girardi}, {Bressan}, {Chiosi}, {Bertelli}  \&
  {Nasi}}{{Girardi} et~al.}{1996}]{gbc+96}
{Girardi} L.,  {Bressan} A.,  {Chiosi} C.,  {Bertelli} G.,   {Nasi} E.,  1996,
  \aaps, \href {http://adsabs.harvard.edu/abs/1996A%26AS..117..113G} {117, 113}

\bibitem[\protect\citeauthoryear{{Gladders} \& {Yee}}{{Gladders} \&
  {Yee}}{2005}]{gy05}
{Gladders} M.~D.,  {Yee} H.~K.~C.,  2005, \mn@doi [\apjs] {10.1086/427327},
  \href {http://adsabs.harvard.edu/abs/2005ApJS..157....1G} {157, 1}

\bibitem[\protect\citeauthoryear{{Gonzalez} et~al.,}{{Gonzalez}
  et~al.}{2019}]{ggb+19}
{Gonzalez} A.~H.,  et~al., 2019, \mn@doi [\apjs] {10.3847/1538-4365/aafad2},
  \href {https://ui.adsabs.harvard.edu/abs/2019ApJS..240...33G} {240, 33}

\bibitem[\protect\citeauthoryear{{Goto} et~al.,}{{Goto} et~al.}{2002}]{gsn+02}
{Goto} T.,  et~al., 2002, \mn@doi [\aj] {10.1086/339303}, \href
  {http://adsabs.harvard.edu/abs/2002AJ....123.1807G} {123, 1807}

\bibitem[\protect\citeauthoryear{{Grove}, {Benoist}  \& {Martel}}{{Grove}
  et~al.}{2009}]{gbm09}
{Grove} L.~F.,  {Benoist} C.,   {Martel} F.,  2009, \mn@doi [\aap]
  {10.1051/0004-6361:200810384}, \href
  {http://adsabs.harvard.edu/abs/2009A%26A...494..845G} {494, 845}

\bibitem[\protect\citeauthoryear{{Hao} et~al.,}{{Hao} et~al.}{2010}]{hmk+10}
{Hao} J.,  et~al., 2010, \mn@doi [\apjs] {10.1088/0067-0049/191/2/254}, \href
  {http://adsabs.harvard.edu/abs/2010ApJS..191..254H} {191, 254}

\bibitem[\protect\citeauthoryear{{Hasinger} et~al.,}{{Hasinger}
  et~al.}{2007}]{hcb+07}
{Hasinger} G.,  et~al., 2007, \mn@doi [\apjs] {10.1086/516576}, \href
  {https://ui.adsabs.harvard.edu/abs/2007ApJS..172...29H} {172, 29}

\bibitem[\protect\citeauthoryear{{Hasselfield} et~al.,}{{Hasselfield}
  et~al.}{2013}]{hhm+13}
{Hasselfield} M.,  et~al., 2013, \mn@doi [\jcap]
  {10.1088/1475-7516/2013/07/008}, \href
  {http://adsabs.harvard.edu/abs/2013JCAP...07..008H} {7, 008}

\bibitem[\protect\citeauthoryear{{Hilton} et~al.,}{{Hilton}
  et~al.}{2018}]{hhs+18}
{Hilton} M.,  et~al., 2018, \mn@doi [\apjs] {10.3847/1538-4365/aaa6cb}, \href
  {http://adsabs.harvard.edu/abs/2018ApJS..235...20H} {235, 20}

\bibitem[\protect\citeauthoryear{{Hoyle}, {Jimenez}  \& {Verde}}{{Hoyle}
  et~al.}{2011}]{hjv11}
{Hoyle} B.,  {Jimenez} R.,   {Verde} L.,  2011, \mn@doi [\prd]
  {10.1103/PhysRevD.83.103502}, \href
  {https://ui.adsabs.harvard.edu/abs/2011PhRvD..83j3502H} {83, 103502}

\bibitem[\protect\citeauthoryear{{Hsieh} \& {Yee}}{{Hsieh} \&
  {Yee}}{2014}]{hy14}
{Hsieh} B.~C.,  {Yee} H.~K.~C.,  2014, \mn@doi [\apj]
  {10.1088/0004-637X/792/2/102}, \href
  {https://ui.adsabs.harvard.edu/abs/2014ApJ...792..102H} {792, 102}

\bibitem[\protect\citeauthoryear{{Ilbert} et~al.,}{{Ilbert}
  et~al.}{2006}]{iam+06}
{Ilbert} O.,  et~al., 2006, \mn@doi [\aap] {10.1051/0004-6361:20065138}, \href
  {http://adsabs.harvard.edu/abs/2006A%26A...457..841I} {457, 841}

\bibitem[\protect\citeauthoryear{{Jarrett} et~al.,}{{Jarrett}
  et~al.}{2011}]{jcm+11}
{Jarrett} T.~H.,  et~al., 2011, \mn@doi [\apj] {10.1088/0004-637X/735/2/112},
  \href {http://adsabs.harvard.edu/abs/2011ApJ...735..112J} {735, 112}

\bibitem[\protect\citeauthoryear{{Jee} et~al.,}{{Jee} et~al.}{2009}]{jrf+09}
{Jee} M.~J.,  et~al., 2009, \mn@doi [\apj] {10.1088/0004-637X/704/1/672}, \href
  {https://ui.adsabs.harvard.edu/abs/2009ApJ...704..672J} {704, 672}

\bibitem[\protect\citeauthoryear{{Kale}, {Venturi}, {Cassano}, {Giacintucci},
  {Bardelli}, {Dallacasa}  \& {Zucca}}{{Kale} et~al.}{2015}]{kvc+15}
{Kale} R.,  {Venturi} T.,  {Cassano} R.,  {Giacintucci} S.,  {Bardelli} S.,
  {Dallacasa} D.,   {Zucca} E.,  2015, \mn@doi [\aap]
  {10.1051/0004-6361/201526341}, \href
  {https://ui.adsabs.harvard.edu/abs/2015A&A...581A..23K} {581, A23}

\bibitem[\protect\citeauthoryear{{Kashino} et~al.,}{{Kashino}
  et~al.}{2019}]{kss+19}
{Kashino} D.,  et~al., 2019, \mn@doi [\apjs] {10.3847/1538-4365/ab06c4}, \href
  {https://ui.adsabs.harvard.edu/abs/2019ApJS..241...10K} {241, 10}

\bibitem[\protect\citeauthoryear{{Knobel} et~al.,}{{Knobel}
  et~al.}{2012}]{kli+12}
{Knobel} C.,  et~al., 2012, \mn@doi [\apj] {10.1088/0004-637X/753/2/121}, \href
  {https://ui.adsabs.harvard.edu/abs/2012ApJ...753..121K} {753, 121}

\bibitem[\protect\citeauthoryear{{Koester} et~al.,}{{Koester}
  et~al.}{2007}]{kma+07b}
{Koester} B.~P.,  et~al., 2007, \mn@doi [\apj] {10.1086/509599}, \href
  {http://ads.ari.uni-heidelberg.de/abs/2007ApJ...660..239K} {660, 239}

\bibitem[\protect\citeauthoryear{{Kriek}, {van Dokkum}, {Labb{\'e}}, {Franx},
  {Illingworth}, {Marchesini}  \& {Quadri}}{{Kriek} et~al.}{2009}]{kvl+09}
{Kriek} M.,  {van Dokkum} P.~G.,  {Labb{\'e}} I.,  {Franx} M.,  {Illingworth}
  G.~D.,  {Marchesini} D.,   {Quadri} R.~F.,  2009, \mn@doi [\apj]
  {10.1088/0004-637X/700/1/221}, \href
  {https://ui.adsabs.harvard.edu/abs/2009ApJ...700..221K} {700, 221}

\bibitem[\protect\citeauthoryear{{Laigle} et~al.,}{{Laigle}
  et~al.}{2016}]{lmi+16}
{Laigle} C.,  et~al., 2016, \mn@doi [\apjs] {10.3847/0067-0049/224/2/24}, \href
  {https://ui.adsabs.harvard.edu/abs/2016ApJS..224...24L} {224, 24}

\bibitem[\protect\citeauthoryear{{Laird} et~al.,}{{Laird}
  et~al.}{2009}]{lng+09}
{Laird} E.~S.,  et~al., 2009, \mn@doi [\apjs] {10.1088/0067-0049/180/1/102},
  \href {https://ui.adsabs.harvard.edu/abs/2009ApJS..180..102L} {180, 102}

\bibitem[\protect\citeauthoryear{{Lang}}{{Lang}}{2014}]{lang14}
{Lang} D.,  2014, \mn@doi [\aj] {10.1088/0004-6256/147/5/108}, \href
  {https://ui.adsabs.harvard.edu/abs/2014AJ....147..108L} {147, 108}

\bibitem[\protect\citeauthoryear{{Le F{\`e}vre} et~al.,}{{Le F{\`e}vre}
  et~al.}{2013}]{lcc+13}
{Le F{\`e}vre} O.,  et~al., 2013, \mn@doi [\aap] {10.1051/0004-6361/201322179},
  \href {https://ui.adsabs.harvard.edu/abs/2013A%26A...559A..14L} {559, A14}

\bibitem[\protect\citeauthoryear{{Li}, {Wu}, {Cao}  \& {Zhu}}{{Li}
  et~al.}{2007}]{lwc+07}
{Li} H.-N.,  {Wu} H.,  {Cao} C.,   {Zhu} Y.-N.,  2007, \mn@doi [\aj]
  {10.1086/520807}, \href
  {https://ui.adsabs.harvard.edu/abs/2007AJ....134.1315L} {134, 1315}

\bibitem[\protect\citeauthoryear{{Lidman} et~al.,}{{Lidman}
  et~al.}{2012}]{lsm+12}
{Lidman} C.,  et~al., 2012, \mn@doi [\mnras]
  {10.1111/j.1365-2966.2012.21984.x}, \href
  {http://adsabs.harvard.edu/abs/2012MNRAS.427..550L} {427, 550}

\bibitem[\protect\citeauthoryear{{Lilly} et~al.,}{{Lilly}
  et~al.}{2009}]{llm+09}
{Lilly} S.~J.,  et~al., 2009, \mn@doi [\apjs] {10.1088/0067-0049/184/2/218},
  \href {https://ui.adsabs.harvard.edu/abs/2009ApJS..184..218L} {184, 218}

\bibitem[\protect\citeauthoryear{{Lima}, {Cunha}, {Oyaizu}, {Frieman}, {Lin}
  \& {Sheldon}}{{Lima} et~al.}{2008}]{lco+08}
{Lima} M.,  {Cunha} C.~E.,  {Oyaizu} H.,  {Frieman} J.,  {Lin} H.,   {Sheldon}
  E.~S.,  2008, \mn@doi [\mnras] {10.1111/j.1365-2966.2008.13510.x}, \href
  {https://ui.adsabs.harvard.edu/abs/2008MNRAS.390..118L} {390, 118}

\bibitem[\protect\citeauthoryear{{Lin} et~al.,}{{Lin} et~al.}{2017}]{lhl+17}
{Lin} Y.-T.,  et~al., 2017, \mn@doi [\apj] {10.3847/1538-4357/aa9bf5}, \href
  {http://adsabs.harvard.edu/abs/2017ApJ...851..139L} {851, 139}

\bibitem[\protect\citeauthoryear{{Liske} et~al.,}{{Liske}
  et~al.}{2015}]{lbd+15}
{Liske} J.,  et~al., 2015, \mn@doi [\mnras] {10.1093/mnras/stv1436}, \href
  {https://ui.adsabs.harvard.edu/abs/2015MNRAS.452.2087L} {452, 2087}

\bibitem[\protect\citeauthoryear{{Liu}, {Mao}  \& {Meng}}{{Liu}
  et~al.}{2012}]{lmm12}
{Liu} F.~S.,  {Mao} S.,   {Meng} X.~M.,  2012, \mn@doi [\mnras]
  {10.1111/j.1365-2966.2012.20886.x}, \href
  {http://adsabs.harvard.edu/abs/2012MNRAS.423..422L} {423, 422}

\bibitem[\protect\citeauthoryear{{Mainzer} et~al.,}{{Mainzer}
  et~al.}{2014}]{mbc+14}
{Mainzer} A.,  et~al., 2014, \mn@doi [\apj] {10.1088/0004-637X/792/1/30}, \href
  {https://ui.adsabs.harvard.edu/abs/2014ApJ...792...30M} {792, 30}

\bibitem[\protect\citeauthoryear{{Mancone}, {Gonzalez}, {Brodwin}, {Stanford},
  {Eisenhardt}, {Stern}  \& {Jones}}{{Mancone} et~al.}{2010}]{mgb+10}
{Mancone} C.~L.,  {Gonzalez} A.~H.,  {Brodwin} M.,  {Stanford} S.~A.,
  {Eisenhardt} P.~R.~M.,  {Stern} D.,   {Jones} C.,  2010, \mn@doi [\apj]
  {10.1088/0004-637X/720/1/284}, \href
  {http://adsabs.harvard.edu/abs/2010ApJ...720..284M} {720, 284}

\bibitem[\protect\citeauthoryear{{Mantz}, {Allen}, {Ebeling}, {Rapetti}  \&
  {Drlica-Wagner}}{{Mantz} et~al.}{2010}]{mae+10}
{Mantz} A.,  {Allen} S.~W.,  {Ebeling} H.,  {Rapetti} D.,   {Drlica-Wagner} A.,
   2010, \mn@doi [\mnras] {10.1111/j.1365-2966.2010.16993.x}, \href
  {http://adsabs.harvard.edu/abs/2010MNRAS.406.1773M} {406, 1773}

\bibitem[\protect\citeauthoryear{{McLure} et~al.,}{{McLure}
  et~al.}{2013}]{mpd+13}
{McLure} R.~J.,  et~al., 2013, \mn@doi [\mnras] {10.1093/mnras/sts092}, \href
  {https://ui.adsabs.harvard.edu/abs/2013MNRAS.428.1088M} {428, 1088}

\bibitem[\protect\citeauthoryear{{Momcheva} et~al.,}{{Momcheva}
  et~al.}{2016}]{mbv+16}
{Momcheva} I.~G.,  et~al., 2016, \mn@doi [\apjs] {10.3847/0067-0049/225/2/27},
  \href {https://ui.adsabs.harvard.edu/abs/2016ApJS..225...27M} {225, 27}

\bibitem[\protect\citeauthoryear{{Nishizawa}, {Hsieh}, {Tanaka}  \&
  {Takata}}{{Nishizawa} et~al.}{2020}]{nht+20}
{Nishizawa} A.~J.,  {Hsieh} B.-C.,  {Tanaka} M.,   {Takata} T.,  2020, arXiv
  e-prints, \href {https://ui.adsabs.harvard.edu/abs/2020arXiv200301511N} {p.
  arXiv:2003.01511}

\bibitem[\protect\citeauthoryear{{Oguri}}{{Oguri}}{2014}]{ogu14}
{Oguri} M.,  2014, \mn@doi [\mnras] {10.1093/mnras/stu1446}, \href
  {http://adsabs.harvard.edu/abs/2014MNRAS.444..147O} {444, 147}

\bibitem[\protect\citeauthoryear{{Oguri} et~al.,}{{Oguri}
  et~al.}{2018}]{oll+18}
{Oguri} M.,  et~al., 2018, \mn@doi [\pasj] {10.1093/pasj/psx042}, \href
  {http://adsabs.harvard.edu/abs/2018PASJ...70S..20O} {70, S20}

\bibitem[\protect\citeauthoryear{{Pacaud} et~al.,}{{Pacaud}
  et~al.}{2016}]{pcg+16}
{Pacaud} F.,  et~al., 2016, \mn@doi [\aap] {10.1051/0004-6361/201526891}, \href
  {http://adsabs.harvard.edu/abs/2016A%26A...592A...2P} {592, A2}

\bibitem[\protect\citeauthoryear{{Papovich} et~al.,}{{Papovich}
  et~al.}{2010}]{pmw+10}
{Papovich} C.,  et~al., 2010, \mn@doi [\apj] {10.1088/0004-637X/716/2/1503},
  \href {http://adsabs.harvard.edu/abs/2010ApJ...716.1503P} {716, 1503}

\bibitem[\protect\citeauthoryear{{Peebles}}{{Peebles}}{1980}]{pee80}
{Peebles} P.~J.~E.,  1980, {The large-scale structure of the universe}

\bibitem[\protect\citeauthoryear{{Pierre} et~al.,}{{Pierre}
  et~al.}{2004}]{pva+04}
{Pierre} M.,  et~al., 2004, \mn@doi [\jcap] {10.1088/1475-7516/2004/09/011},
  \href {https://ui.adsabs.harvard.edu/abs/2004JCAP...09..011P} {9, 011}

\bibitem[\protect\citeauthoryear{{Pierre} et~al.,}{{Pierre}
  et~al.}{2016}]{ppa+16}
{Pierre} M.,  et~al., 2016, \mn@doi [\aap] {10.1051/0004-6361/201526766}, \href
  {https://ui.adsabs.harvard.edu/abs/2016A%26A...592A...1P} {592, A1}

\bibitem[\protect\citeauthoryear{{Piffaretti}, {Arnaud}, {Pratt},
  {Pointecouteau}  \& {Melin}}{{Piffaretti} et~al.}{2011}]{pap+11}
{Piffaretti} R.,  {Arnaud} M.,  {Pratt} G.~W.,  {Pointecouteau} E.,   {Melin}
  J.-B.,  2011, \mn@doi [\aap] {10.1051/0004-6361/201015377}, \href
  {http://adsabs.harvard.edu/abs/2011A%26A...534A.109P} {534, A109}

\bibitem[\protect\citeauthoryear{{Planck Collaboration} et~al.,}{{Planck
  Collaboration} et~al.}{2016}]{plancksz16}
{Planck Collaboration} et~al., 2016, \mn@doi [\aap]
  {10.1051/0004-6361/201525823}, \href
  {http://adsabs.harvard.edu/abs/2016A%26A...594A..27P} {594, A27}

\bibitem[\protect\citeauthoryear{{Prakash} et~al.,}{{Prakash}
  et~al.}{2016}]{pln+16}
{Prakash} A.,  et~al., 2016, \mn@doi [\apjs] {10.3847/0067-0049/224/2/34},
  \href {http://adsabs.harvard.edu/abs/2016ApJS..224...34P} {224, 34}

\bibitem[\protect\citeauthoryear{{Reichardt} et~al.,}{{Reichardt}
  et~al.}{2013}]{rsb+13}
{Reichardt} C.~L.,  et~al., 2013, \mn@doi [\apj] {10.1088/0004-637X/763/2/127},
  \href {https://ui.adsabs.harvard.edu/abs/2013ApJ...763..127R} {763, 127}

\bibitem[\protect\citeauthoryear{{Rigopoulou} et~al.,}{{Rigopoulou}
  et~al.}{2006}]{rhp+06}
{Rigopoulou} D.,  et~al., 2006, \mn@doi [\apj] {10.1086/505784}, \href
  {https://ui.adsabs.harvard.edu/abs/2006ApJ...648...81R} {648, 81}

\bibitem[\protect\citeauthoryear{{Rood}}{{Rood}}{1969}]{roo69}
{Rood} H.~J.,  1969, \mn@doi [\apj] {10.1086/150227}, \href
  {http://adsabs.harvard.edu/abs/1969ApJ...158..657R} {158, 657}

\bibitem[\protect\citeauthoryear{{Rykoff} et~al.,}{{Rykoff}
  et~al.}{2014}]{rrb+14}
{Rykoff} E.~S.,  et~al., 2014, \mn@doi [\apj] {10.1088/0004-637X/785/2/104},
  \href {http://adsabs.harvard.edu/abs/2014ApJ...785..104R} {785, 104}

\bibitem[\protect\citeauthoryear{{Sanders} et~al.,}{{Sanders}
  et~al.}{2007}]{ssa+07}
{Sanders} D.~B.,  et~al., 2007, \mn@doi [\apjs] {10.1086/517885}, \href
  {https://ui.adsabs.harvard.edu/abs/2007ApJS..172...86S} {172, 86}

\bibitem[\protect\citeauthoryear{{Schlafly}, {Meisner}  \& {Green}}{{Schlafly}
  et~al.}{2019}]{smg19}
{Schlafly} E.~F.,  {Meisner} A.~M.,   {Green} G.~M.,  2019, \mn@doi [\apjs]
  {10.3847/1538-4365/aafbea}, \href
  {https://ui.adsabs.harvard.edu/abs/2019ApJS..240...30S} {240, 30}

\bibitem[\protect\citeauthoryear{{Scoville} et~al.,}{{Scoville}
  et~al.}{2007a}]{sab+07}
{Scoville} N.,  et~al., 2007a, \mn@doi [\apjs] {10.1086/516585}, \href
  {https://ui.adsabs.harvard.edu/abs/2007ApJS..172....1S} {172, 1}

\bibitem[\protect\citeauthoryear{{Scoville} et~al.,}{{Scoville}
  et~al.}{2007b}]{saa+07}
{Scoville} N.,  et~al., 2007b, \mn@doi [\apjs] {10.1086/516580}, \href
  {https://ui.adsabs.harvard.edu/abs/2007ApJS..172...38S} {172, 38}

\bibitem[\protect\citeauthoryear{{Shu}, {Koposov}, {Evans}, {Belokurov},
  {McMahon}, {Auger}  \& {Lemon}}{{Shu} et~al.}{2019}]{ske+19}
{Shu} Y.,  {Koposov} S.~E.,  {Evans} N.~W.,  {Belokurov} V.,  {McMahon} R.~G.,
  {Auger} M.~W.,   {Lemon} C.~A.,  2019, \mn@doi [\mnras]
  {10.1093/mnras/stz2487}, \href
  {https://ui.adsabs.harvard.edu/abs/2019MNRAS.489.4741S} {489, 4741}

\bibitem[\protect\citeauthoryear{{Silverman} et~al.,}{{Silverman}
  et~al.}{2015}]{sks+15}
{Silverman} J.~D.,  et~al., 2015, \mn@doi [\apjs] {10.1088/0067-0049/220/1/12},
  \href {https://ui.adsabs.harvard.edu/abs/2015ApJS..220...12S} {220, 12}

\bibitem[\protect\citeauthoryear{{Skelton} et~al.,}{{Skelton}
  et~al.}{2014}]{swm+14}
{Skelton} R.~E.,  et~al., 2014, \mn@doi [\apjs] {10.1088/0067-0049/214/2/24},
  \href {https://ui.adsabs.harvard.edu/abs/2014ApJS..214...24S} {214, 24}

\bibitem[\protect\citeauthoryear{{Smith}, {Gear}, {Smith}, {Papageorgiou}  \&
  {Eales}}{{Smith} et~al.}{2019}]{sgs+19}
{Smith} C.~M.~A.,  {Gear} W.~K.,  {Smith} M.~W.~L.,  {Papageorgiou} A.,
  {Eales} S.~A.,  2019, \mn@doi [\mnras] {10.1093/mnras/stz1090}, \href
  {https://ui.adsabs.harvard.edu/abs/2019MNRAS.486.4304S} {486, 4304}

\bibitem[\protect\citeauthoryear{{S{\"o}chting}, {Coldwell}, {Clowes},
  {Campusano}  \& {Graham}}{{S{\"o}chting} et~al.}{2012}]{scc+12}
{S{\"o}chting} I.~K.,  {Coldwell} G.~V.,  {Clowes} R.~G.,  {Campusano} L.~E.,
  {Graham} M.~J.,  2012, \mn@doi [\mnras] {10.1111/j.1365-2966.2012.21050.x},
  \href {https://ui.adsabs.harvard.edu/abs/2012MNRAS.423.2436S} {423, 2436}

\bibitem[\protect\citeauthoryear{{Stott}, {Edge}, {Smith}, {Swinbank}  \&
  {Ebeling}}{{Stott} et~al.}{2008}]{ses+08}
{Stott} J.~P.,  {Edge} A.~C.,  {Smith} G.~P.,  {Swinbank} A.~M.,   {Ebeling}
  H.,  2008, \mn@doi [\mnras] {10.1111/j.1365-2966.2007.12807.x}, \href
  {http://adsabs.harvard.edu/abs/2008MNRAS.384.1502S} {384, 1502}

\bibitem[\protect\citeauthoryear{{Szabo}, {Pierpaoli}, {Dong}, {Pipino}  \&
  {Gunn}}{{Szabo} et~al.}{2011}]{spd+11}
{Szabo} T.,  {Pierpaoli} E.,  {Dong} F.,  {Pipino} A.,   {Gunn} J.,  2011,
  \mn@doi [\apj] {10.1088/0004-637X/736/1/21}, \href
  {http://adsabs.harvard.edu/abs/2011ApJ...736...21S} {736, 21}

\bibitem[\protect\citeauthoryear{{Tanaka}}{{Tanaka}}{2015}]{tan15}
{Tanaka} M.,  2015, \mn@doi [\apj] {10.1088/0004-637X/801/1/20}, \href
  {https://ui.adsabs.harvard.edu/abs/2015ApJ...801...20T} {801, 20}

\bibitem[\protect\citeauthoryear{{Tanaka} et~al.,}{{Tanaka}
  et~al.}{2018}]{tch+18}
{Tanaka} M.,  et~al., 2018, \mn@doi [\pasj] {10.1093/pasj/psx077}, \href
  {https://ui.adsabs.harvard.edu/abs/2018PASJ...70S...9T} {70, S9}

\bibitem[\protect\citeauthoryear{{Taniguchi} et~al.,}{{Taniguchi}
  et~al.}{2007}]{tsm+07}
{Taniguchi} Y.,  et~al., 2007, \mn@doi [\apjs] {10.1086/516596}, \href
  {https://ui.adsabs.harvard.edu/abs/2007ApJS..172....9T} {172, 9}

\bibitem[\protect\citeauthoryear{{Tarr{\'\i}o} \& {Zarattini}}{{Tarr{\'\i}o} \&
  {Zarattini}}{2020}]{tz20}
{Tarr{\'\i}o} P.,  {Zarattini} S.,  2020, arXiv e-prints, \href
  {https://ui.adsabs.harvard.edu/abs/2020arXiv200506489T} {p. arXiv:2005.06489}

\bibitem[\protect\citeauthoryear{{Tinker}, {Kravtsov}, {Klypin}, {Abazajian},
  {Warren}, {Yepes}, {Gottl{\"o}ber}  \& {Holz}}{{Tinker}
  et~al.}{2008}]{tkk+08}
{Tinker} J.,  {Kravtsov} A.~V.,  {Klypin} A.,  {Abazajian} K.,  {Warren} M.,
  {Yepes} G.,  {Gottl{\"o}ber} S.,   {Holz} D.~E.,  2008, \mn@doi [\apj]
  {10.1086/591439}, \href {http://adsabs.harvard.edu/abs/2008ApJ...688..709T}
  {688, 709}

\bibitem[\protect\citeauthoryear{{Tozzi} et~al.,}{{Tozzi} et~al.}{2015}]{tsj15}
{Tozzi} P.,  et~al., 2015, \mn@doi [\apj] {10.1088/0004-637X/799/1/93}, \href
  {https://ui.adsabs.harvard.edu/abs/2015ApJ...799...93T} {799, 93}

\bibitem[\protect\citeauthoryear{{Vikhlinin} et~al.,}{{Vikhlinin}
  et~al.}{2009}]{vkb+09}
{Vikhlinin} A.,  et~al., 2009, \mn@doi [\apj] {10.1088/0004-637X/692/2/1060},
  \href {http://adsabs.harvard.edu/abs/2009ApJ...692.1060V} {692, 1060}

\bibitem[\protect\citeauthoryear{{Wang} et~al.,}{{Wang} et~al.}{2016}]{wed+16}
{Wang} T.,  et~al., 2016, \mn@doi [\apj] {10.3847/0004-637X/828/1/56}, \href
  {https://ui.adsabs.harvard.edu/abs/2016ApJ...828...56W} {828, 56}

\bibitem[\protect\citeauthoryear{{Webb} et~al.,}{{Webb} et~al.}{2015}]{wmn+15}
{Webb} T. M.~A.,  et~al., 2015, \mn@doi [\apj] {10.1088/0004-637X/814/2/96},
  \href {https://ui.adsabs.harvard.edu/abs/2015ApJ...814...96W} {814, 96}

\bibitem[\protect\citeauthoryear{{Wen} \& {Han}}{{Wen} \& {Han}}{2011}]{wh11}
{Wen} Z.~L.,  {Han} J.~L.,  2011, \mn@doi [\apj] {10.1088/0004-637X/734/1/68},
  \href {http://adsabs.harvard.edu/abs/2011ApJ...734...68W} {734, 68}

\bibitem[\protect\citeauthoryear{{Wen} \& {Han}}{{Wen} \& {Han}}{2015}]{wh15}
{Wen} Z.~L.,  {Han} J.~L.,  2015, \mn@doi [\apj] {10.1088/0004-637X/807/2/178},
  \href {http://adsabs.harvard.edu/abs/2015ApJ...807..178W} {807, 178}

\bibitem[\protect\citeauthoryear{{Wen} \& {Han}}{{Wen} \& {Han}}{2018}]{wh18}
{Wen} Z.~L.,  {Han} J.~L.,  2018, \mn@doi [\mnras] {10.1093/mnras/sty2533},
  \href {https://ui.adsabs.harvard.edu/abs/2018MNRAS.481.4158W} {481, 4158}

\bibitem[\protect\citeauthoryear{{Wen}, {Han}  \& {Liu}}{{Wen}
  et~al.}{2009}]{whl09}
{Wen} Z.~L.,  {Han} J.~L.,   {Liu} F.~S.,  2009, \mn@doi [\apjs]
  {10.1088/0067-0049/183/2/197}, \href
  {http://adsabs.harvard.edu/abs/2009ApJS..183..197W} {183, 197}

\bibitem[\protect\citeauthoryear{{Wen}, {Han}  \& {Liu}}{{Wen}
  et~al.}{2012}]{whl12}
{Wen} Z.~L.,  {Han} J.~L.,   {Liu} F.~S.,  2012, \mn@doi [\apjs]
  {10.1088/0067-0049/199/2/34}, \href
  {http://adsabs.harvard.edu/abs/2012ApJS..199...34W} {199, 34}

\bibitem[\protect\citeauthoryear{{Wen}, {Wu}, {Zhu}, {Lam}, {Wu}, {Wicker}  \&
  {Zhao}}{{Wen} et~al.}{2013}]{wwz+13}
{Wen} X.-Q.,  {Wu} H.,  {Zhu} Y.-N.,  {Lam} M.~I.,  {Wu} C.-J.,  {Wicker} J.,
  {Zhao} Y.-H.,  2013, \mn@doi [\mnras] {10.1093/mnras/stt939}, \href
  {https://ui.adsabs.harvard.edu/abs/2013MNRAS.433.2946W} {433, 2946}

\bibitem[\protect\citeauthoryear{{Wen}, {Han}  \& {Yang}}{{Wen}
  et~al.}{2018}]{why18}
{Wen} Z.~L.,  {Han} J.~L.,   {Yang} F.,  2018, \mn@doi [\mnras]
  {10.1093/mnras/stx3189}, \href
  {http://adsabs.harvard.edu/abs/2018MNRAS.475..343W} {475, 343}

\bibitem[\protect\citeauthoryear{{Westera}, {Lejeune}, {Buser}, {Cuisinier}  \&
  {Bruzual}}{{Westera} et~al.}{2002}]{wlb+02}
{Westera} P.,  {Lejeune} T.,  {Buser} R.,  {Cuisinier} F.,   {Bruzual} G.,
  2002, \mn@doi [\aap] {10.1051/0004-6361:20011493}, \href
  {http://cdsads.u-strasbg.fr/abs/2002A%26A...381..524W} {381, 524}

\bibitem[\protect\citeauthoryear{{Whiley} et~al.,}{{Whiley}
  et~al.}{2008}]{wad+08}
{Whiley} I.~M.,  et~al., 2008, \mn@doi [\mnras]
  {10.1111/j.1365-2966.2008.13324.x}, \href
  {http://adsabs.harvard.edu/abs/2008MNRAS.387.1253W} {387, 1253}

\bibitem[\protect\citeauthoryear{{Willis}, {Ramos-Ceja}, {Muzzin}, {Pacaud},
  {Yee}  \& {Wilson}}{{Willis} et~al.}{2018}]{wrm+18}
{Willis} J.~P.,  {Ramos-Ceja} M.~E.,  {Muzzin} A.,  {Pacaud} F.,  {Yee}
  H.~K.~C.,   {Wilson} G.,  2018, \mn@doi [\mnras] {10.1093/mnras/sty975},
  \href {https://ui.adsabs.harvard.edu/abs/2018MNRAS.477.5517W} {477, 5517}

\bibitem[\protect\citeauthoryear{{Wright} et~al.,}{{Wright}
  et~al.}{2010}]{wem+10}
{Wright} E.~L.,  et~al., 2010, \mn@doi [\aj] {10.1088/0004-6256/140/6/1868},
  \href {http://adsabs.harvard.edu/abs/2010AJ....140.1868W} {140, 1868}

\bibitem[\protect\citeauthoryear{{Zamojski} et~al.,}{{Zamojski}
  et~al.}{2007}]{zsr+07}
{Zamojski} M.~A.,  et~al., 2007, \mn@doi [\apjs] {10.1086/516593}, \href
  {https://ui.adsabs.harvard.edu/abs/2007ApJS..172..468Z} {172, 468}

\bibitem[\protect\citeauthoryear{{Zatloukal}, {R{\"o}ser}, {Wolf}, {Hippelein}
  \& {Falter}}{{Zatloukal} et~al.}{2007}]{zrw+07}
{Zatloukal} M.,  {R{\"o}ser} H.-J.,  {Wolf} C.,  {Hippelein} H.,   {Falter} S.,
   2007, \mn@doi [\aap] {10.1051/0004-6361:20078063}, \href
  {http://adsabs.harvard.edu/abs/2007A%26A...474L...5Z} {474, L5}

\bibitem[\protect\citeauthoryear{{Zhang} et~al.,}{{Zhang}
  et~al.}{2016}]{zmm+16}
{Zhang} Y.,  et~al., 2016, \mn@doi [\apj] {10.3847/0004-637X/816/2/98}, \href
  {https://ui.adsabs.harvard.edu/abs/2016ApJ...816...98Z} {816, 98}

\bibitem[\protect\citeauthoryear{{Zou}, {Gao}, {Zhou}  \& {Kong}}{{Zou}
  et~al.}{2019}]{zgz+19}
{Zou} H.,  {Gao} J.,  {Zhou} X.,   {Kong} X.,  2019, \mn@doi [\apjs]
  {10.3847/1538-4365/ab1847}, \href
  {https://ui.adsabs.harvard.edu/abs/2019ApJS..242....8Z} {242, 8}

\bibitem[\protect\citeauthoryear{{van Breukelen} et~al.,}{{van Breukelen}
  et~al.}{2006}]{vcb+06}
{van Breukelen} C.,  et~al., 2006, \mn@doi [\mnras]
  {10.1111/j.1745-3933.2006.00236.x}, \href
  {http://adsabs.harvard.edu/abs/2006MNRAS.373L..26V} {373, L26}

\bibitem[\protect\citeauthoryear{{van der Burg}, {McGee}, {Aussel}, {Dahle},
  {Arnaud}, {Pratt}  \& {Muzzin}}{{van der Burg} et~al.}{2018}]{vrm+18}
{van der Burg} R. F.~J.,  {McGee} S.,  {Aussel} H.,  {Dahle} H.,  {Arnaud} M.,
  {Pratt} G.~W.,   {Muzzin} A.,  2018, \mn@doi [\aap]
  {10.1051/0004-6361/201833572}, \href
  {https://ui.adsabs.harvard.edu/abs/2018A&A...618A.140V} {618, A140}

\makeatother
\end{thebibliography}

\label{lastpage}
\end{document}